\documentstyle[12pt,preprint,aps,floats,epsf]{revtex}
\tighten

\begin{document}

\preprint{
\noindent
\hfill
\begin{minipage}[t]{6in}
\begin{flushright}
CERN-TH/99-205  \\
SISSA 121/99/EP \\
BUTP--99/14     \\
MPI/PhT-99-37\\
CALT-68-2239\\
ZH-TH 24/99\\
hep-ph/9911245  \\
\vspace*{1.0cm}
\end{flushright}
\end{minipage}
}

\draft

\title { Gluino Contribution to Radiative B Decays:\\[1.2ex]
 Organization of QCD Corrections and Leading Order Results}

\author{Francesca Borzumati$^{\,a,b}$, Christoph Greub$^{\,c}$,
        Tobias Hurth$^{\,a,d}$, and Daniel Wyler$^{\,e}$ }  
\address{$^{a}$
  Theory Division, CERN, CH--1211 Geneva 23, Switzerland}
\address{$^{b}$
Scuola Internazionale Superiore di Studi Avanzati (SISSA),
 I--34013 Trieste, Italy}
\address{$^{c}$ 
 Institut f\"ur Theoretische Physik, Universit\"at Bern, 
 CH--3012 Bern, Switzerland} 
\address{$^{d}$  
 Max-Planck-Institut f\"ur Physik, Werner-Heisenberg-Institut,
 D--80805 Munich, Germany}  
\address{$^{e}$
 Institut f\"ur Theoretische Physik, Universit\"at Z\"urich,
 CH--8057 Zurich, Switzerland} 

\maketitle

\vspace*{1truecm}
\begin{abstract}
The gluino-induced contributions to the decay $b \to s \gamma$ are
investigated in supersymmetric frameworks with generic sources of
flavour violation.  It is shown that, when QCD corrections are taken
into account, the relevant operator basis of the Standard Model
effective Hamiltonian gets enlarged to contain:
i) magnetic and chromomagnetic operators with a factor of
 $\alpha_s$ and weighted by a quark mass $m_b$ or $m_c$;
ii) magnetic and chromomagnetic operators of lower dimensionality,
 also containing $\alpha_s$; 
iii) four-quark operators weighted by a factor $\alpha_s^2$.  
Numerical results are given, showing the effects of the leading order
QCD corrections on the inclusive branching ratio for 
$b \to s \gamma$. Constraints on supersymmetric sources of flavour
violation are derived.
\end{abstract}

\vfill
October 1999 


\setlength{\parskip}{1.2ex}

\section{Introduction}
\label{intro}

Processes involving Flavour Changing Neutral Currents (FCNCs) provide
invaluable guidelines for supersymmetric model building. The
experimental measurements of the rates for these processes, or the
upper limits set on them, impose in general a reduction of the large
number and size of parameters in the soft supersymmetry-breaking terms
present in these models. Among these processes, those involving
transitions between first- and second-generation quarks, namely FCNC
processes in the $K$ system, are considered as the most formidable
tools to shape viable supersymmetric flavour models. Moreover, the
tight experimental bounds on some flavour-diagonal transitions, such
as the electric dipole moment of the electron and of the neutron, as
well as $g-2$, help constraining soft terms inducing chirality
violations.

Several supersymmetric models have so far emerged, with specific
solutions to the chiral-flavour problem.  Among them are two classes
of models in which the dynamics of flavour sets in above the
supersymmetry breaking scale and in which the subsequent flavour
problem is killed by the mechanisms of communicating supersymmetry
breaking to the experimentally accessible sector. They are known as
mSUGRA, i.e. minimal supersymmetric standard models in which
supergravity is the mediator between the supersymmetry-breaking sector
and the visible sector~\cite{MSUGRA}, and gauge-mediated
supersymmetry-breaking models (GMSBs)~\cite{GMSB}, in which the
communication between the two sectors is realized by gauge
interactions.  In other classes of models, particular flavour
symmetries are introduced, which link quarks and squarks: models in
which an alignment of squarks and quarks is assumed~\cite{ALIGNEMENT},
and models in which the solution to the flavour problem is obtained by
advocating heavy first- and second-generation
squarks~\cite{HEAVY1,HEAVY2-UTWO,HEAVY3,UTWO}. In the latter, the
splitting between squarks of first and second generation and those
belonging to the third generation relies on a $U(2)$ flavour
symmetry~\cite{HEAVY2-UTWO,UTWO}.

Neutral flavour transitions involving third-generation quarks do not
yet pose serious threats to these models. One exception comes from the
decay $b\to s\gamma$, the least rare flavour- and chirality-violating
process in the $B$ system. It has been detected, but the precision of
the experimental measurement of its rate is not very high at the
moment. Nevertheless, this measurement already has the effect of
carving out some regions in the space of free parameters of most of
the models in the above classes (see for example~\cite{BDN}; for a
recent analysis, see~\cite{OKADA} and references therein). They also
drastically constrain several somewhat tuned realizations of models in
these classes~\cite{BOP,MMM}.  Once the precision in the experimental
measurement has increased, this decay will undoubtedly gain efficiency
in selecting the viable regions of the parameter space in the above
classes of models and it may help discriminating among the models by
then proposed.  It is therefore important to get ready reliable
calculations of this decay rate, i.e. calculations in which
theoretical uncertainties are reduced as much as possible, and which
are general enough to be applied to generic supersymmetric models.

The experimental situation is, at present, as follows.  The ALEPH
Collaboration at LEP reports a value of the inclusive decay 
$\bar{B}\to X_s\gamma$ of~\cite{ALEPH}:
\begin{equation}
 {\rm BR}(\bar{B}\to X_s\gamma) =
  (3.11\pm 0.80\pm0.72)\times 10^{-4}
\label{alephres}
\end{equation}
from a sample of $b$ hadrons at the $Z$ resonance. The CLEO
Collaboration at CESR has a statistically and systematically more
precise result, based on $3.3\times 10^{6}$ $B \bar{B}$
events~\cite{CLEO}:
\begin{equation}
 {\rm BR}(\bar{B}\to X_s\gamma)
  = (3.15 \pm 0.35\pm 0.32 \pm 0.26)\times 10^{-4}\,,
\label{cleores}
\end{equation}
but quotes a still very large interval~\cite{CLEO},  
\begin{equation}
    2 \times 10^{-4} 
  <  {\rm BR}(\bar{B}\to X_s\gamma)
  < 4.5\times 10^{-4},  
\label{cleoband}
\end{equation}
as the range of acceptable values of branching ratios.

Theoretically, the rate for this decay, characterized by its
large QCD contributions, practically as large as the purely
electroweak ones~\cite{BBM+DLTES}, is known with high accuracy in the
Standard Model (SM). It has been calculated up to the next-to-leading
order (NLO) in QCD, using the formalism of effective
Hamiltonians~\cite{EFFHAM}. 
Results for LO and NLO calculations and for power corrections can be
found
in~\cite{HISTORY0,HISTORY1,MISIAK95},~\cite{ALI,CMM,HISTORY2,CHW,BG},
and~\cite{HISTORY3}, respectively.  The resulting theoretical accuracy
is rather astonishing: the inclusion of the NLO QCD corrections
reduces the large scale dependences that are present at LO 
($\pm 25\%$) to
a mere per cent uncertainty, once the value of the parameters to be
input in this calculation is fixed. This accuracy, however, is
obtained through large and accidental numerical cancellations among
different contributions to the NLO corrections and a subsequent
cancellation of scale dependences~\cite{BG,KN}. The same accuracy,
indeed, is not obtained for the NLO calculation of the rate 
${\rm BR}(\bar{B}\to X_s\gamma)$ in simple extensions of the SM, such 
as models that differ from the SM by the addition of two or more
doublets to the Higgs sector~\cite{BG}.

The calculation of ${\rm BR}(\bar{B}\to X_s\gamma)$ within
supersymmetric models is still far from this level of
sophistication. There are several contributions to the amplitude of
this decay, usually identified by the particles exchanged in the
loop. Besides the $W^-$--$t$-quark and $H^-$--$t$-quark contributions,
there are also the chargino, gluino and neutralino contributions,
respectively mediated by the exchange of chargino--up-squarks,
gluino--down-squarks and neutralino--down-squarks. All these
contributions were calculated in Ref.~\cite{BBMR} within mSUGRA;
their analytic expressions apply naturally to GMSB models also.
The inclusion of QCD corrections needed for the calculation of the
rate, was assumed in~\cite{BBMR} to follow the SM pattern.  No
dedicated study of this decay exists for the supersymmetric models
mentioned above with specific flavour symmetries. A calculation of
${\rm BR}(\bar{B}\to X_s\gamma)$ induced solely by the gluino
contribution has been performed in~\cite{BBM0,GGMS} for a generic
supersymmetric model, but no QCD corrections were included.

A NLO analysis of ${\rm BR}(\bar{B}\to X_s\gamma)$ 
was recently performed~\cite{CDGGsusy} for a specific supersymmetric 
case (the corresponding 
NLO matching conditions are also given in~\cite{BMU}). 
This is valid in a 
class of models where the only source of flavour violation at the
electroweak scale is that of the SM, encoded in the
Cabibbo--Kobayashi--Maskawa (CKM) matrix. It applies to
mSUGRA and GMSB models (in which the same features are
assumed/obtained at the messenger scale) only when the amount of
flavour violation, generated radiatively between the
supersymmetry-breaking scale and the electroweak scale, can be
neglected with respect to that induced by the CKM matrix. It applies,
therefore, to the case in which only the lightest stop eigenstate
contributes to the chargino-mediated loop and all other squarks and 
gluino are heavy enough to be decoupled at the electroweak scale. It
cannot be used in particular directions of parameter space of the
above listed models in which quantum effects induce a gluino
contribution~\cite{DNW} as large as the chargino or the SM
contribution~\cite{MMM,FB}.  Nor can it be used as a
model-discriminator tool, able to constrain the potentially large
sources of flavour violation typical of generic supersymmetric models.

Among these, flavour-violating scalar mass terms and trilinear terms
induce a flavour non-diagonal vertex gluino--quark--squark.  This is
generically assumed to provide the dominant contributions to quark-flavour 
transitions thanks to its large coupling $g_s$. Therefore, it
is often taken as the only contribution to these
transitions~\cite{HKT}, and in particular to the $b \to s \gamma $
decay, when attempting to obtain order-of-magnitude upper bounds on
flavour-violating terms in the scalar potential~\cite{BBM0,GGMS}. Once
the constraints coming from experimental measurements are imposed,
however, the gluino contribution is reduced to values such that the SM
and the other supersymmetric contributions can no longer  be neglected.  
Any LO and NLO calculation of the $b \to s \gamma$ rate in
generic supersymmetric models should then include all possible
contributions.

The gluino contribution presents some peculiar features,
related to the implementation of QCD corrections, that have not 
been detected so
far. As already mentioned, the decay $b \to s \gamma$ involves a
quark-flavour violation as well as chirality violation.  The first is
directly related to the flavour violation in the virtual sfermions
exchanged in the loop. The second can be obtained as in the SM,
through a chirality flip in the external $b$-quark, and it is
signalled by its mass $\overline{m_b}$. It can also be induced by
sfermion mass terms originating from trilinear soft
supersymmetry-breaking terms. These mass terms differ from fermionic
mass terms by two units of $R$-charge under a $U(1)_R$ symmetry. The
correct $R$-charge for this $b$--$s$ transition is then restored
through the insertion of the gluino mass $m_{\widetilde g}$ in the
gluino propagator.  The two different mechanisms producing chirality
violation are well known. They give rise to operators of different
dimensionality when generating the effective Hamiltonian used to
include QCD corrections to the $b \to s \gamma $ decay. Indeed,
$m_{\widetilde g}$, the mass of one of the heavy fields exchanged in
the loop, is naturally incorporated in the Wilson coefficient of the
corresponding magnetic operator, which is now of dimension $five$
($ e \,g_s^2 \,
 (\bar{s} \sigma^{\mu\nu} P_R b) \, F_{\mu\nu}$). On the contrary,
$\overline{m_b}$, the running mass of one light field, with
a full dynamics below the matching scale, is naturally included in the
definition of a magnetic operator, which is of dimension $six$
($ e \,g_s^2 \,{\overline m}_b \,
 (\bar{s} \sigma^{\mu\nu} P_R b) \, F_{\mu\nu}$).

Moreover, the presence of the strong coupling $\alpha_s$ in the gluino
contribution immediately sparks off the question of whether this coupling
should be included in the definition of the gluino-induced operators
or in the corresponding Wilson coefficients. 
Both choices are, in principle, acceptable.
It can be observed, however, as will be discussed
in Sec.~\ref{ordersplit}, that the first option does not
require a modification of the program of implementation of QCD
corrections established in the SM case. In particular, 
the anomalous dimension matrix starts at order $\alpha_s$ and 
is used up to order $\alpha_s$ 
($\alpha_s^2$) in a LO (NLO) calculation.
The inclusion of the
$\alpha_s$ coupling in the operators imposes a necessary
distinction of the dimension {\it six} gluino-induced magnetic
operators
$ e \,g_s^2 \,{\overline m}_b \,
 (\bar{s} \sigma^{\mu\nu} P_R b) \, F_{\mu\nu}$
from the SM magnetic operator
$  {e/16\pi^2} \,{\overline m}_b \,
 (\bar{s} \sigma^{\mu\nu} P_R b) \, F_{\mu\nu}$. 
As it will be seen in Sec.~\ref{ordersplit},
a set of new four-fermion operators, induced by gluino exchanges, 
is also needed.

These features single out the gluino contribution to the decay 
$b \to s \gamma$ as one that necessarily requires a dedicated study of
the implementation of QCD corrections already at the LO in QCD, before
including chargino and neutralino contributions and higher-order QCD
corrections. In Sec.~\ref{ordersplit}, the list of operators induced
by gluino-mediated loops is given together with the list of those
needed for the SM contribution.  The number of operators depends on
the sources of flavour violation that are present in the particular
supersymmetric model considered. In the attempt to reach the level of
generality advocated above, no restriction is made on the possible
sources of flavour violation in the sfermion sector. These are
surveyed in Sec.~\ref{flavchange}. Also shown is the direct connection
between flavour-violating sources and operators generated, emphasizing
the differences between the analysis in a generic supersymmetric model
and the typical mSUGRA-inspired analyses. The Wilson coefficients at
the matching scale for the Hamiltonian generated by gluino
contributions are given in Sec.~\ref{WilsonCoeff}.  They are
calculated using the mass-eigenstate formalism, the most appropriate
to deal with different off-diagonal terms in the sfermion mass matrix
squared, of a priori unknown size.  These coefficients evolve down to
the low-scale $\mu_b$ independently of the usual SM coefficients,
since there is no mixing between SM and gluino-induced operators.  The
anomalous-dimension matrix governing this evolution at the LO in QCD
and the resulting analytic expressions for the low-scale Wilson
coefficients is given in Sec.~\ref{gmatrix}.  In 
Sec.~\ref{branching},
an expression for the LO rate ${\rm BR}(\bar{B}\to X_s\gamma)$, due to
the SM and the gluino-induced Wilson coefficients, is derived.
Numerical evaluations of the branching ratio are shown in
Sec.~\ref{constraints}, when only one or at most two off-diagonal
elements in the down-squark mass matrix squared are non-vanishing.  As
already mentioned, the decay $b \to s \gamma$ can be realistically
used as a tool to select viable supersymmetric flavour models only
when all contributions to ${\rm BR}(\bar{B}\to X_s\gamma)$ are
included. The numerical evaluations of Sec.~\ref{constraints},
therefore, have only the purpose of illustrating the effect of the LO
QCD corrections, as well as the interplay between SM and gluino
contributions to the branching ratio. Strictly speaking, they give
results that are valid only in particular directions of the parameter
space of generic supersymmetric models, and provide, in general, some
intermediate results of an ongoing, more complete analysis.

\section{Ordering the QCD Perturbative Expansion and the
  effective Hamiltonian}
\label{ordersplit}

In the SM, rare $B$-meson decays are induced by loops in which $W$
bosons and up-type quarks propagate.  The most important corrections
are due to exchanges of light particles, gluons and light quarks,
which give rise to powers of the large logarithmic factor
$L=\log(m_b^2/m_W^2)$.

The decay amplitude for $b \to s \gamma$ obtains large logarithms $L$
only from loops with gluons.  This implies at least one factor of
$\alpha_s$ for each large logarithm.  Since the two scales $m_b$ and
$M_W$ are far apart, $L$ is a large number and these terms need to be
resummed: powers of $\alpha_s L$ are resummed at the LO, terms of the
form $\alpha_s \, (\alpha_s L)^N$ are obtained at the NLO.  Thus, the
corrections to the decay amplitude are classified according to:
\begin{itemize}
 \item[] \quad (LO): 
 $\quad \quad G_F \, (\alpha_s L)^N, \quad \quad (N=0,1,...)$  
\item[] \quad (NLO):
 $\quad       G_F \, \alpha_s (\alpha_s L)^N$,
\end{itemize}
where $G_F$ is the Fermi constant. 

The resummation of these corrections is usually achieved by making use of
the formalism of effective Hamiltonians, combined with renormalization
group techniques. The needed effective Hamiltonian is obtained by 
integrating out the heavy degrees of freedom, i.e.  the top-quark and
the $W$ boson. It is usually expressed as
\begin{equation}
 {\cal H}_{eff}^{W} = 
 - \frac{4 G_F}{\sqrt{2}} V_{tb}^{\phantom{\ast}} V_{ts}^\ast
  \sum_i C_i(\mu) {\cal O}_i(\mu) \,, 
\label{weffham}
\end{equation}
where $V_{tb}$ and $V_{ts}$ are
elements of the Cabibbo--Kobayashi--Maskawa (CKM) matrix. The Wilson
coefficients $C_i$ contain all dependence on the heavy degrees of
freedom, whereas the operators ${\cal O}_{i}$ depend on light
fields only. The operators relevant to radiative $B$ decays can be 
divided into two classes:
\\[1.02ex] $\bullet$ \ 
current--current operators and gluonic penguin operators~\cite{MISIAK95}:
\begin{equation}
\begin{array}{llll}
{\cal O}_{1}                 \,= &\!
(\bar{s} \gamma_\mu T^a P_L c)\,  (\bar{c} \gamma^\mu T_a P_L b)\,, 
               &         \\[1.2ex]
{\cal O}_{2}                 \,= &\!
(\bar{s} \gamma_\mu P_L c)\,  (\bar{c} \gamma^\mu P_L b)\,,    
               &         \\[1.2ex]
{\cal O}_{3}                 \,= &\!
(\bar{s} \gamma_\mu P_L b) \sum_q (\bar{q} \gamma^\mu q)\,,    
               &        \\[1.2ex]
{\cal O}_{4}                 \,= &\!
(\bar{s} \gamma_\mu T^a P_L b) \sum_q (\bar{q} \gamma^\mu T_a q)\,,
               &        \\[1.2ex]
{\cal O}_{5}                 \,= &\!
(\bar{s} \gamma_\mu \gamma_\nu \gamma_\rho P_L b) 
 \sum_q (\bar{q} \gamma^\mu \gamma^\nu \gamma^\rho q)\,,        
               &        \\[1.2ex]
{\cal O}_{6}                 \,= &\!
(\bar{s} \gamma_\mu \gamma_\nu \gamma_\rho T^a P_L b) 
 \sum_q (\bar{q} \gamma^\mu \gamma^\nu \gamma^\rho T_a q)\,,      
               & 
\end{array}
\label{smfourquarkop}
\end{equation}
where $T^a$ ($a=1,8$) are $SU(3)$ colour generators; 
\\[1.02ex] $\bullet$ \ 
magnetic operators, with chirality violation signalled by the 
presence of the $b$-quark mass:
\begin{equation}
\begin{array}{ll}
{\cal O}_{7}                 \,= &\!
  \displaystyle{\frac{e}{16\pi^2}} \,{\overline m}_b(\mu) \,
 (\bar{s} \sigma^{\mu\nu} P_R b) \, F_{\mu\nu}\,,      \\[2.0ex]    
{\cal O}_{8}                 \,= &\!
  \displaystyle{\frac{g_s}{16\pi^2}} \,{\overline m}_b(\mu) \,
 (\bar{s} \sigma^{\mu\nu} T^a P_R b)
     \, G^a_{\mu\nu}\,,                               
\label{smmagnop}                                        
\end{array}
\end{equation}
where $g_s$ and $e$ are the strong and electromagnetic coupling 
constants.  
Both sets of operators, those in~(\ref{smfourquarkop}) and 
in~(\ref{smmagnop}) are of dimension $six$. 

It is by now well known that a consistent calculation for 
$b \to s \gamma$ at LO (or NLO) precision requires three steps:
\begin{itemize}
\item[{\it 1)}] 
a matching calculation of the full standard model theory 
with the effective theory at the scale $\mu=\mu_W$ 
to order $\alpha_s^0$ (or $\alpha_s^1$) for Wilson coefficients, 
where  $\mu_W$ denotes a scale of order $M_W$ or $m_t$;
\item[{\it 2)}]  
a renormalization group treatment of the Wilson coefficients
using the anomalous-dimension matrix to order $\alpha_s^1$ 
(or $\alpha_s^2$);
\item[{\it 3)}]   
a calculation of the operator matrix elements at the scale 
$\mu = \mu_b$  to order $\alpha_s^0$ (or $\alpha_s^1$), 
where $\mu_b$ denotes a scale of order $m_b$.
\end{itemize}

That matters can be somewhat different is illustrated by
the decay $b \to s \, \ell \, \bar{\ell}$. The effective
Hamiltonian~(\ref{weffham}) contains in this case two additional
operators:
\begin{equation}
\begin{array}{ll}
{\cal O}_{9\phantom{a}}                 \,= &\!
  \displaystyle{\frac{e^2}{16\pi^2}} \,
 (\bar{s} \gamma_\mu  P_L b)\, (\bar{\ell} \gamma^\mu \ell)\,, \\[2.0ex]
{\cal O}_{10}                           \,= &\! 
  \displaystyle{\frac{e^2}{16\pi^2}} \,
 (\bar{s} \gamma_\mu P_L b)\,  (\bar{\ell} \gamma^\mu \gamma_5 \ell)\,.    
\end{array}
\label{leptonop}
\end{equation}
It turns out that in this case, 
the operator ${\cal O}_2$ mixes into ${\cal O}_9$ at one loop:
the pair $c \bar{c}$ in ${\cal O}_2$ can be closed to form a loop,
and an off-shell photon producing a pair $\ell \,\bar{\ell}$ can 
be radiated from a quark line. 
The first large logarithm $L=\log(m^2_b/M^2_W)$ arises without the 
exchange of gluons. This possibility has no 
correspondence in the $b\to s \gamma$ case. 
Consequently, the decay amplitude is ordered according to 
$ G_F L \, (\alpha_s L)^N$ at the LO in QCD and 
$ G_F \alpha_s L (\alpha_s L)^N$ at the NLO.
To achieve technically the resummation of these terms, it is convenient
to  redefine magnetic, chromomagnetic and lepton-pair operators 
$ {\cal O}_{7}$, $ {\cal O}_{8}$, $ {\cal O}_{9}$, 
and $ {\cal O}_{10}$ and the corresponding coefficients 
as follows~\cite{M-BM}:
\begin{equation}
\label{reshuffle}
{\cal O}_i^{new} = \frac{16 \pi^2}{g_s^2} {\cal O}_i\,, 
\quad \quad 
C_i^{new} = \frac{g_s^2}{16\pi^2} C_i \quad 
\quad (i=7,...,10). 
\end{equation}
This redefinition allows us to proceed according to 
the above three steps when calculating the 
amplitude of the decay $b \to s \, \ell \, \bar{\ell}$~\cite{M-BM}.
In particular, the one-loop mixing of the
operator ${\cal O}_2$ with the operator ${\cal O}_9^{new}$ appears
formally at ${\cal O}(\alpha_s)$.



In supersymmetric models, where the gluino--quark--squark vertex can be 
flavour violating, the exchange of gluino and squarks in the loop
gives contribution to the decay $b \rightarrow s \gamma$. 
Various combinations of the gluino--quark--squark vertex lead to
$|\Delta(B)|=|\Delta(S)|=1$ magnetic and chromomagnetic
operators (of ${\cal O}_7$-type, ${\cal O}_8$-type) with an explicit 
factor $\alpha_s$, and to four-quark operators, with a factor 
$\alpha_s^2$.  The
complete effective Hamiltonian can then be split in two terms:
\begin{equation} 
{\cal H}_{eff} = {\cal H}_{eff}^W + {\cal H}_{eff}^{\tilde{g}} \,,
\label{hfull}
\end{equation} 
where $ {\cal H}_{eff}^{W}$ is the SM effective Hamiltonian
in~(\ref{weffham}) and ${\cal H}_{eff}^{\tilde{g}}$ originates after
integrating out squarks and gluinos. 
Note that `mixed' diagrams, which contain, besides a $W$ boson, also
gluinos and squarks, give rise to $\alpha_s$ corrections to the Wilson
coefficients in ${\cal H}_{eff}^W$ (at the matching scale). Such
contributions can be omitted in a LO calculation, but they have to be
taken into account at the NLO level.

As far as the gluino-induced contribution to the decay amplitude
$ b \to s \gamma$ is concerned, the aim is to resum
the following terms:
\begin{itemize}
\item[] \quad (LO):
 $\quad \quad \alpha_s \, (\alpha_s L)^N, \quad \quad (N=0,1,...)$  
\item[] \quad (NLO): 
 $\quad \alpha_s \, \alpha_s (\alpha_s L)^N$,  
\end{itemize}
respectively at the leading and next-to-leading order.  

\begin{figure}[t]
\begin{center}
\leavevmode
\epsfxsize= 6.0 truecm
\epsfbox[200 560 425 680]{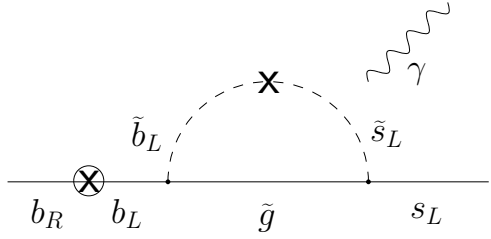}
\end{center}
\caption[f1]{Diagram mediating the $b \to s \gamma$ decay through
 gluino exchange and contributing to the operator 
 ${\cal O}_{7b,\tilde{g}}$. A contribution to the primed operator 
 ${\cal O}_{7b,\tilde{g}}^{\prime}$ is obtained by exchanging
 $L\leftrightarrow R$.} 
\label{glopb}
\end{figure}
While ${\cal H}_{eff}^{\tilde{g}}$ is unambiguous, it is a matter of
convention whether the $\alpha_s$ factors, peculiar of the gluino 
exchange, should be put into the
definition of operators or into the Wilson coefficients.  In analogy
to the decay $b \to s \ell^+ \ell^- $ discussed above, it is
convenient to distribute the factors of $\alpha_s$ between operators
and Wilson coefficients in such a way that the first two of the three
steps in the program for the SM calculation also apply to the
gluino-induced contribution.  This implies one factor of $\alpha_s^1$
in the definition of the magnetic and chromomagnetic operators and a
factor $\alpha_s^2$ in the definition of the four-quark operators.
With this convention, the matching calculation and the evolution down to
the low scale $\mu_b$ of the Wilson coefficients are organized exactly
in the same way as in the SM.  The anomalous-dimension matrix, indeed,
has the canonical expansion in $\alpha_s$ and starts with a term
proportional to $\alpha_s^1$.  The last of the three steps in the program 
of the SM calculation 
requires now an obvious modification: the
calculation of the matrix elements has to be performed at order
$\alpha_s$ and $\alpha_s^2$ at the LO and NLO precision.  With this
organization of QCD corrections, the SM Hamiltonian 
${\cal H}_{eff}^{W}$ in eq.~(\ref{weffham}) and the gluino-induced one 
${\cal H}_{eff}^{\tilde{g}}$ undergo separate renormalization, which
facilitates all considerations.

\begin{figure}[t]
\begin{center}
\leavevmode
\epsfxsize= 6.0 truecm
\epsfbox[200 560 425 680]{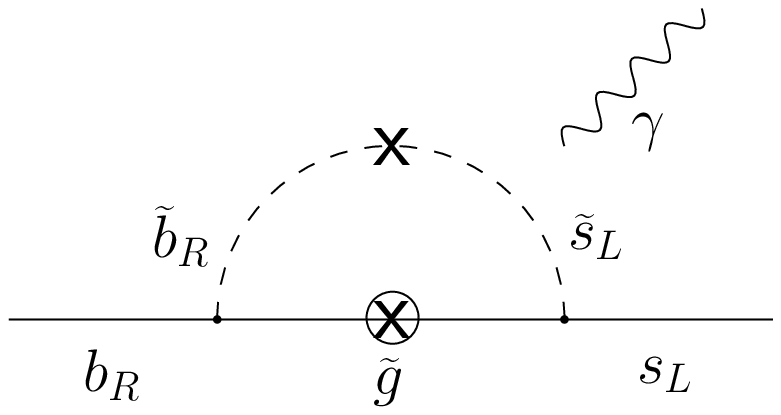}
\end{center}
\caption[f1]{Contribution to ${\cal O}_{7\tilde{g},\tilde{g}}$ from
 the insertion of the gluino mass and of a scalar mass term
 simultaneously violating chirality and flavour. A contribution to
 ${\cal O}_{7\tilde{g},\tilde{g}}^{\prime}$ is obtained through the
 interchange $L\leftrightarrow R$.}
\label{glopgoneins}
\vspace*{0.3truecm}
\begin{center}
\leavevmode
\epsfxsize= 6.0 truecm
\epsfbox[200 560 425 680]{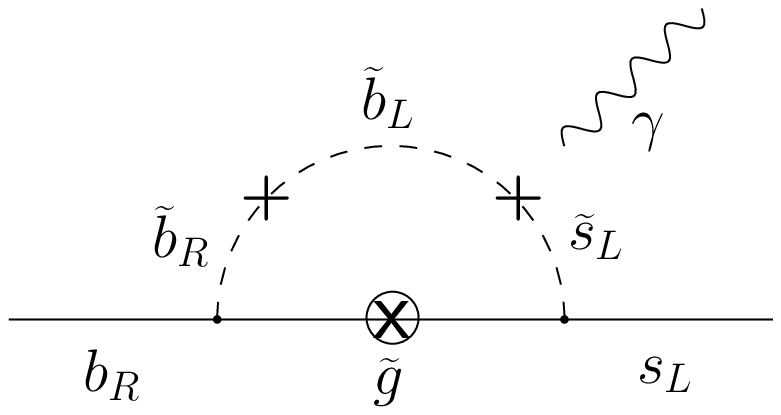}
\hspace*{1.0truecm}
\epsfxsize= 6.0 truecm
\epsfbox[200 560 425 680]{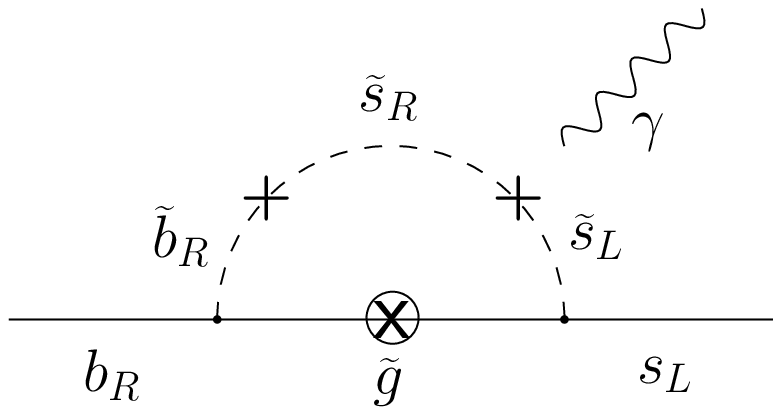}
\end{center}
\caption[f1]{Contributions to ${\cal O}_{7\tilde{g},\tilde{g}}$ from 
 the insertion of the gluino mass and distinct chirality- and
 flavour-violating scalar mass terms. In the approximation $m_s=0$,
 the second diagram requires trilinear terms not linked to Yukawa
 couplings.  The analogous contributions to
 ${\cal O}_{7\tilde{g},\tilde{g}}^{\prime}$ are obtained through the
 interchange $L \leftrightarrow R$.}
\label{glopgtwoins}
\end{figure}

\begin{figure}[t]
\begin{center}
\leavevmode
\epsfxsize= 6.0 truecm
\epsfbox[200 510 425 640]{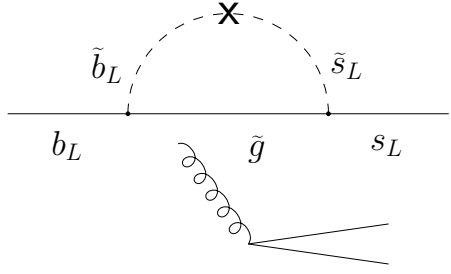}
\end{center}
\caption[f1]{Penguin diagram contributing to the 
 operators~(\ref{penguinboxop}).}
\label{penguindiagr}
\end{figure}

\begin{figure}[t]
\begin{center}
\leavevmode
\epsfxsize= 6.0 truecm
\epsfbox[200 560 425 680]{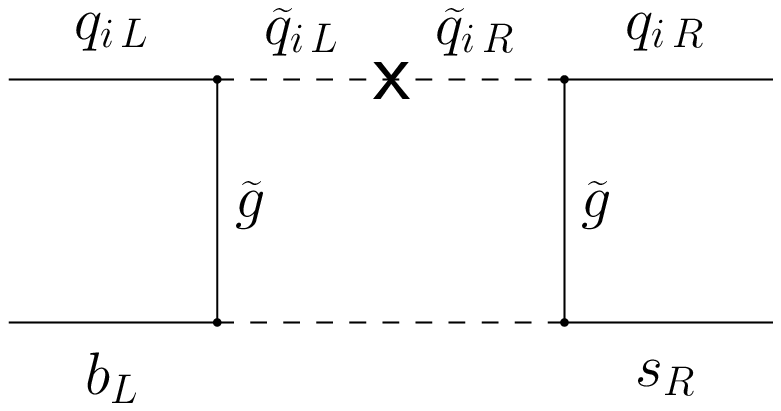}
\hspace*{1.0truecm}
\epsfxsize= 6.0 truecm
\epsfbox[200 560 420 680]{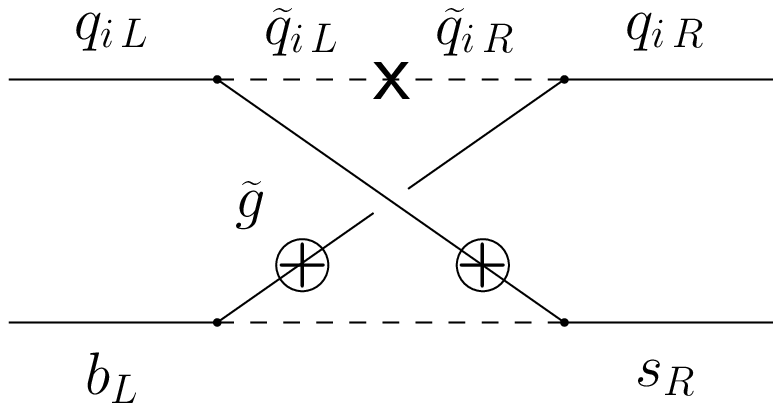}
\end{center}
\begin{center}
\leavevmode
\epsfxsize= 6.0 truecm
\epsfbox[200 560 425 680]{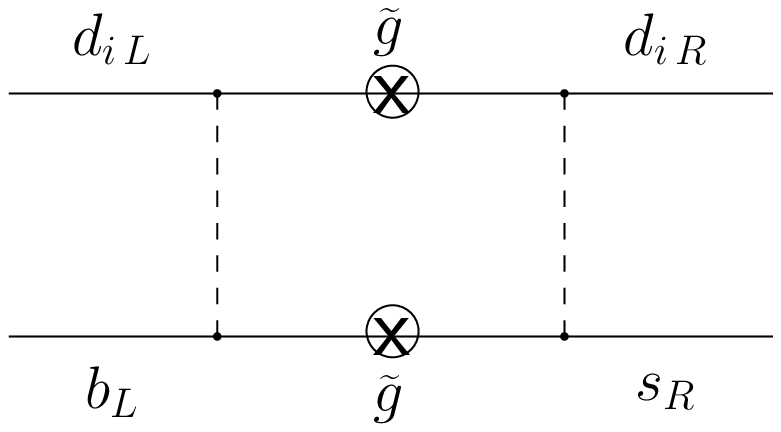}
\hspace*{1.0truecm}
\epsfxsize= 6.0 truecm
\epsfbox[200 560 420 680]{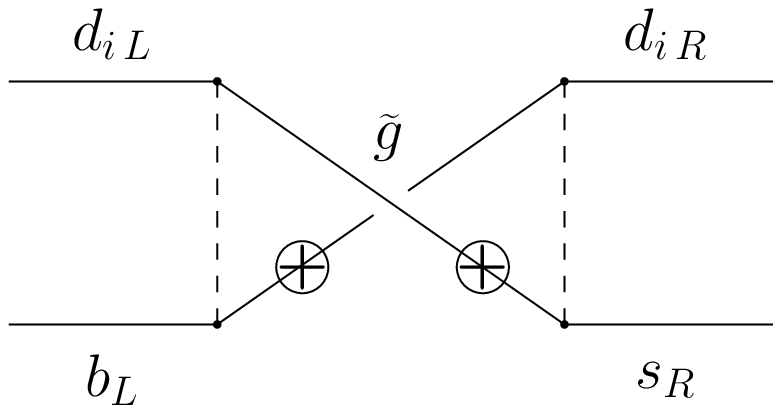}
\end{center}
\caption[f1]{Diagrams contributing to the
 operators~(\ref{boxop}).  In the two upper
 diagrams, the quark (squark) $q$ ($\widetilde{q}$) can be of up- or
 down-type and the flavour violation on the lower squark line, not
 explicitly indicated, can be realized through a direct flavour--chiral
 transition (see Fig.~\ref{glopgoneins}) or through distinct chirality
 and flavour transitions (see Fig.~\ref{glopgtwoins}). In the lower 
 diagrams, the down-type quark $d_i$ is a $b$- or an $s$-quark if a
 single flavour violation is allowed in the squark lines.}
\label{boxdiagr}
\end{figure}

The effective Hamiltonian ${\cal H}_{eff}^{\tilde{g}}$,
is further split into two parts:
\begin{equation}
 {\cal H}_{eff}^{\tilde{g}} = 
 \sum_i C_{i,\tilde{g}}(\mu) {\cal O}_{i,\tilde{g}}(\mu)  +
 \sum_i \sum_q C_{i,\tilde{g}}^q(\mu) {\cal O}_{i,\tilde{g}}^q(\mu) \,, 
\label{geffham}
\end{equation}
where the index $q$ runs over all light quarks
$q=u,d,c,s,b$. The operators contributing to the first part are:
\\[1.02ex] 
$\bullet$ \ 
magnetic operators, with chirality violation coming 
from the $b$-quark mass:
\begin{equation}
\begin{array}{llll}
{\cal O}_{7b,\tilde{g}}                 \,= &\!
   e \,g_s^2(\mu) \,{\overline m}_b(\mu) \,
 (\bar{s} \sigma^{\mu\nu} P_R b) \, F_{\mu\nu}\,,   
                                        &  
{\cal O}_{7b,\tilde{g}}^{\prime}        \,= &\!
   e \,g_s^2(\mu) \,{\overline m}_b(\mu) \,
 (\bar{s} \sigma^{\mu\nu} P_L b) \, F_{\mu\nu}\,,     \\[2.0ex]    
{\cal O}_{8b,\tilde{g}}                 \,= &\!
 g_s(\mu) \,g_s^2(\mu) \,{\overline m}_b(\mu) \,
 (\bar{s} \sigma^{\mu\nu} T^a P_R b)
     \, G^a_{\mu\nu}\,,                               
                                        &  
{\cal O}_{8b,\tilde{g}}^\prime          \,= &\!
 g_s(\mu) \,g_s^2(\mu) \,{\overline m}_b(\mu) \,
 (\bar{s} \sigma^{\mu\nu} T^a P_L b)
     \, G^a_{\mu\nu}\,,
\label{gmagnopb} 
\end{array}
\end{equation}
of dimension $six$, as the SM operators. 
A contribution to the magnetic operator ${\cal O}_{7b,\tilde{g}}$ 
is shown in Fig.~\ref{glopb}.  In this and the following diagrams, only
the first in the series of possible insertions of chiral-flavour-violating 
scalar mass terms is drawn.  This has the advantage of
showing pictorially the correlation among supersymmetric sources of
flavour violation and the generation of operators contributing to the
effective Hamiltonian~(\ref{geffham}). Nevertheless, the actual
calculations presented in this paper are performed using squark mass
eigenstates, i.e. resumming over all possible scalar mass insertions.
\\[1.02ex] 
$\bullet$ \ 
magnetic operators in which the chirality-violating parameter is the 
gluino mass $m_{\tilde{g}}$, 
included in the corresponding Wilson coefficients:
\begin{equation}          
\begin{array}{llll}  
{\cal O}_{7\tilde{g},\tilde{g}}         \,= &\!
  e \,g_s^2(\mu) \,
 (\bar{s} \sigma^{\mu\nu} P_R b) \, F_{\mu\nu}\,,    
                                        &  \quad 
{\cal O}_{7\tilde{g},\tilde{g}}^\prime  \,= &\!
  e \,g_s^2(\mu) \,
 (\bar{s} \sigma^{\mu\nu} P_L b) \, F_{\mu\nu}\,,     \\
                                        &             \\[-1.3ex]           
{\cal O}_{8\tilde{g},\tilde{g}}         \,= &\!
 g_s(\mu) \,g_s^2(\mu) \,
 (\bar{s} \sigma^{\mu\nu} T^a P_R b)
     \, G^a_{\mu\nu}\,, 
                                        &  \quad 
{\cal O}_{8\tilde{g},\tilde{g}}^\prime  \,= &\!
 g_s(\mu) \,g_s^2(\mu) \,
 (\bar{s} \sigma^{\mu\nu} T^a P_L b)
     \, G^a_{\mu\nu}\,. 
\label{gmagnopg}                                     
\end{array} 
\end{equation}
Notice that these operators have dimension {\it five}, i.e. 
dimensionality lower than that of all remaining operators, of 
dimension {\it six}. Diagrams generating these operators are shown in 
Figs.~\ref{glopgoneins} and~\ref{glopgtwoins}. 
\\[1.02ex] 
$\bullet$ \ 
magnetic operators, with chirality violation signalled by   
the presence of the $c$-quark mass:
\begin{equation}
\begin{array}{llll}
{\cal O}_{7c,\tilde{g}}                 \,= &\!
   e \,g_s^2(\mu) \,{\overline m}_c(\mu) \,
 (\bar{s} \sigma^{\mu\nu} P_R b) \, F_{\mu\nu}\,,   
                                        &  
{\cal O}_{7c,\tilde{g}}^{\prime}        \,= &\!
   e \,g_s^2(\mu) \,{\overline m}_c(\mu) \,
 (\bar{s} \sigma^{\mu\nu} P_L b) \, F_{\mu\nu}\,,     \\[2.0ex]    
{\cal O}_{8c,\tilde{g}}                 \,= &\!
 g_s(\mu) \,g_s^2(\mu) \,{\overline m}_c(\mu) \,
 (\bar{s} \sigma^{\mu\nu} T^a P_R b)
     \, G^a_{\mu\nu}\,,                               
                                        &  
{\cal O}_{8c,\tilde{g}}^\prime          \,= &\!
 g_s(\mu) \,g_s^2(\mu) \,{\overline m}_c(\mu) \,
 (\bar{s} \sigma^{\mu\nu} T^a P_L b)
     \, G^a_{\mu\nu}\,.             
\label{gmagnopc}                                       
\end{array}
\end{equation}
The origin of these will become clear after discussing the second term
in~(\ref{geffham}). This contains:
\\[1.02ex] 
$\bullet$ \ 
four-quark operators with vector Lorentz structure:
\begin{equation}
\begin{array}{llll}
{\cal O}_{11,\tilde{g}}^q               \,= &\!  g_s^4(\mu) 
(\bar{s} \gamma_\mu  P_L b)\, 
(\bar{q} \gamma^\mu  P_L q) \,,           
               &  \quad \quad
{\cal O}_{11,\tilde{g}}^{q\,\prime}     \,= &\!  g_s^4(\mu)  
(\bar{s} \gamma_\mu  P_R b)\, 
(\bar{q} \gamma^\mu  P_R q) \,,                  \\[1.2ex] 
{\cal O}_{12,\tilde{g}}^q               \,= &\!  g_s^4(\mu) 
(\bar{s}_\alpha \gamma_\mu P_L b_\beta)\,
(\bar{q}_\beta \gamma^\mu P_L q_\alpha) \,,
               &  \quad \quad
{\cal O}_{12,\tilde{g}}^{q\,\prime}     \,= &\!  g_s^4(\mu) 
(\bar{s}_\alpha \gamma_\mu P_R b_\beta)\,
(\bar{q}_\beta \gamma^\mu P_R q_\alpha) \,,      \\[1.2ex]
{\cal O}_{13,\tilde{g}}^q               \,= &\!  g_s^4(\mu) 
(\bar{s} \gamma_\mu P_L b)\,
(\bar{q} \gamma^\mu P_R q) \,,
               &  \quad \quad
{\cal O}_{13,\tilde{g}}^{q\,\prime}     \,= &\!  g_s^4(\mu) 
(\bar{s} \gamma_\mu P_R b)\,
(\bar{q} \gamma^\mu P_L q) \,,                   \\[1.2ex]
{\cal O}_{14,\tilde{g}}^q               \,= &\!  g_s^4(\mu) 
(\bar{s}_\alpha \gamma_\mu P_L b_\beta)\,
(\bar{q}_\beta \gamma^\mu P_R q_\alpha) \,,
               &  \quad \quad
{\cal O}_{14,\tilde{g}}^{q\,\prime}     \,= &\!  g_s^4(\mu) 
(\bar{s}_\alpha \gamma_\mu P_R b_\beta)\,
(\bar{q}_\beta \gamma^\mu P_L q_\alpha) \,,     
\label{penguinboxop}
\end{array}
\end{equation}
where colour indices are omitted for colour-singlet currents.  They
arise from box diagrams through the exchange of two gluinos and from
penguin diagrams through the exchange of a gluino and a gluon. A
typical penguin diagram is shown in Fig.~\ref{penguindiagr}. According to
their Lorentz structure, these
operators will be called hereafter vector four-quark operators.
\\[1.02ex] 
$\bullet$ \ 
four-quark operators with scalar and tensor Lorentz structure:
\begin{equation}
\begin{array}{llll}
{\cal O}_{15,\tilde{g}}^{q}         \,= &\!  g_s^4(\mu) 
(\bar{s} P_R b)\,
(\bar{q} P_R q)\,,
               &  \quad \quad
{\cal O}_{15,\tilde{g}}^{q\,\prime} \,= &\!  g_s^4(\mu) 
(\bar{s} P_L b)\,
(\bar{q} P_L q)\,,                         
               \\[1.2ex]
{\cal O}_{16,\tilde{g}}^{q}         \,= &\!  g_s^4(\mu) 
(\bar{s}_\alpha P_R b_\beta) \,
(\bar{q}_\beta  P_R q_\alpha)\,, 
               &  \quad \quad
{\cal O}_{16,\tilde{g}}^{q\,\prime} \,= &\!  g_s^4(\mu) 
(\bar{s}_\alpha P_L b_\beta) \,
(\bar{q}_\beta  P_L q_\alpha)\,,
               \\[1.2ex]
{\cal O}_{17,\tilde{g}}^{q}         \,= &\!  g_s^4(\mu) 
(\bar{s} P_R b)\,
(\bar{q} P_L q)\,,     
               &  \quad \quad
{\cal O}_{17,\tilde{g}}^{q\,\prime} \,= &\!  g_s^4(\mu) 
(\bar{s} P_L b)\,
(\bar{q} P_R q)\,,             
               \\[1.2ex]
{\cal O}_{18,\tilde{g}}^{q}         \,= &\!  g_s^4(\mu) 
(\bar{s}_\alpha P_R b_\beta) \,
(\bar{q}_\beta  P_L q_\alpha)\,,
               &  \quad \quad
{\cal O}_{18,\tilde{g}}^{q\,\prime} \,= &\!  g_s^4(\mu) 
(\bar{s}_\alpha P_L b_\beta) \,
(\bar{q}_\beta  P_R q_\alpha)\,,           
               \\[1.2ex]
{\cal O}_{19,\tilde{g}}^{q}         \,= &\!  g_s^4(\mu)  
(\bar{s} \sigma_{\mu \nu} P_R b)\,
(\bar{q} \sigma^{\mu \nu} P_R q)\,,
               &  \quad \quad
{\cal O}_{19,\tilde{g}}^{q\,\prime} \,= &\!  g_s^4(\mu) 
(\bar{s} \sigma_{\mu \nu} P_L b)\,
(\bar{q} \sigma^{\mu \nu} P_L q)\,,
               \\[1.2ex]
{\cal O}_{20,\tilde{g}}^{q}         \,= &\!  g_s^4(\mu)  
(\bar{s}_\alpha \sigma_{\mu \nu} P_R b_\beta) \,
(\bar{q}_\beta  \sigma^{\mu \nu} P_R q_\alpha)\,,
               &  \quad \quad
{\cal O}_{20,\tilde{g}}^{q\,\prime} \,= &\!  g_s^4(\mu) 
(\bar{s}_\alpha \sigma_{\mu \nu} P_L b_\beta) \,
(\bar{q}_\beta  \sigma^{\mu \nu} P_L q_\alpha)\,,
\label{boxop}
\end{array} 
\end{equation}
which are induced by box diagrams only and through the exchange of two 
gluinos.  Examples of box diagrams are
sketched in Figs.~\ref{boxdiagr}.  In the following, the
operators~(\ref{boxop}) will be called scalar/tensor four-quark
operators.  Notice that, for different $q$'s,
${\cal O}_{11,\tilde{g}}^q$--${\cal O}_{20,\tilde{g}}^q$ are in
general distinct sets of operators.

The four-quark operators in~(\ref{penguinboxop}) and~(\ref{boxop})
are formally of higher order in the strong coupling than the magnetic
and chromomagnetic operators~(\ref{gmagnopb})--~(\ref{gmagnopc}).  
As it will be explicitly shown in Sec.~\ref{WilsonCoeff}, 
the scalar/tensor operators
${\cal O}_{15,\tilde{g}}^q$--${\cal O}_{20\tilde{g}}^q$ mix at 
one loop into the magnetic and chromomagnetic operators.  
Given this fact, the necessity of including ${\cal O}_{7c,\tilde{g}}$
and ${\cal O}_{8c,\tilde{g}}$ in the operator basis becomes clear
immediately: some of the operators
${\cal O}^q_{15,\tilde{g}}$, ...,
${\cal O}^q_{20,\tilde{g}}$ with $q=c$ mix into 
${\cal O}_{7c,\tilde{g}}$ and
${\cal O}_{8c,\tilde{g}}$. Such mixing terms can be calculated by 
considering the one-loop matrix elements 
$\langle s \gamma|{\cal O}^c_{i,\tilde{g}} |b \rangle$ and
$\langle s g|{\cal O}^c_{i,\tilde{g}} |b \rangle$ $(i=15,...,20)$,
respectively. In principle, also operators like
${\cal O}_{7u,\tilde{g}}$,
${\cal O}_{7d,\tilde{g}}$ and
${\cal O}_{7s,\tilde{g}}$ are induced in an analogous way. These operators, 
however, are weighted by $m_u$, $m_d$ and $m_s$ and are vanishing
in the approximation used here:  $m_u = m_d = m_s = 0$.

Due to these mixing effects,
the scalar/tensor operators have to be included in a LO
calculation for the decay amplitude.
The remaining four-quark
operators with vector structure
${\cal O}_{11,\tilde{g}}^q$--${\cal O}_{14,\tilde{g}}^q$ (and the
corresponding primed operators) do not mix at one loop neither into
the magnetic and chromomagnetic operators nor into the four-quark
operators 
${\cal O}_{15,\tilde{g}}^{q}$--${\cal O}_{20,\tilde{g}}^{q}$.  
Therefore, these vector four-quark operators become relevant only at
the NLO precision.

We end this section with a  comment on the definition of the 
strong coupling constant used in the various steps of the calculation.
In the full theory, which consists here of the SM and 
gluino--down-squark sectors of a supersymmetric model, all particles 
contribute to the running of this coupling,
indicated by the symbol $\hat{g}_s(\mu)$. In order to perform the
matching with the effective theory, where only the five light quarks
survive, all the heavy particles have to be decoupled. The strong
coupling constant in this regime, 
indicated by $g_s(\mu)$, differs from $\hat{g}_s(\mu)$ by
logarithmic terms signalling the decoupling of the heavy particles:
\begin{equation}  
\hat{g}_s(\mu) = g_s(\mu) \, [ 1 + g_s^2(\mu) \mbox{(decoupling log's)}] 
\,.
\end{equation}  
At NLO precision, these decoupling terms have to be taken into account
explicitly. At LO precision, however, $\hat{g}_s(\mu)$ and $g_s(\mu)$
can be identified and $g_s(\mu)$ is here always understood 
to be the $\overline{\mbox{MS}}$ strong coupling at the renormalization 
scale $\mu$, running with five flavours.

\section{Sources of Flavour Violation}
\label{flavchange}

Supersymmetric models contain all sources of flavour violation
present in a Two Higgs Doublet Model of Type~II, i.e. 
the vertices with a charged boson:
$\bar{u}_{L\,i}$--$d_{L\,j}$--$W^+ $ 
and
$\bar{u}_{L\,i}$--$d_{R\,j}$--$H^+ $,
$\bar{u}_{R\,i}$--$d_{L\,j}$--$H^+$ ($i,j=1,2,3$).
Once  
the electroweak symmetry is broken, a rotation in 
flavour space~\cite{HKR} 
\begin{equation}
  D^{\,o}     \  = \ V_d \,D\,,           \hspace*{0.8truecm}
  U^o         \  = \ V_u \,U\,,           \hspace*{0.8truecm}
  D^{\,c\,o}     = \ U_d^\ast \,D^c\,, \hspace*{0.8truecm}
  U^{c\,o}       = \ U_u^\ast \,U^c\,, 
\label{superotation}
\end{equation}
of all matter superfields in the superpotential 
\begin{equation}
 W  \ = \  
  -   D_i^{\,c\,o} \left(h_d\right)_{ij}  Q_j^{\,o} H_d 
  +   U_i^{c\,o}   \left(h_u\right)_{ij}  Q_j^{\,o} H_u
  - \mu H_d H_u  \,,
\label{superpot}
\end{equation}
brings fermions from the current eigenstate basis 
$\{d_L^o,u_L^o,d_R^o,u_R^o\}$ to their mass eigenstate basis 
$\{d_L,u_L,d_R,u_R\}$:  
\begin{equation}
d_L^o = V_d d_L\,, \hspace*{0.8truecm}
u_L^o = V_u u_L\,, \hspace*{0.8truecm}
d_R^o = U_d d_R\,, \hspace*{0.8truecm}
u_R^o = U_u u_R\,, 
\label{fermrotation}
\end{equation}
and the scalar superpartners to the basis 
$\{ \widetilde{D}, \widetilde{U}, 
    \widetilde{D}^{c}, \widetilde{U}^{c} \}$.
Through this rotation, the Yukawa matrices $h_d$ and $h_u$ 
are reduced to their diagonal form $\hat{h}_d$ and $\hat{h}_u$: 
\begin{equation}
(\hat{h}_d)_{ii} = (U_d^\dagger h_d V_d)_{ii}= \frac{m_{d\,i}}{v_d}\,,
\hspace*{1.0truecm} 
(\hat{h}_u)_{ii} = (U_u^\dagger h_u V_u)_{ii}= \frac{m_{u\,i}}{v_u}\,.
\label{diagyukawa}
\end{equation}
Tree-level mixing terms among quarks of different generations are due 
to the misalignment of $V_d$ and $V_u$; all the above vertices 
$\bar{u}_{L\,i}$--$d_{L\,j}$--$W^+ $ and
$\bar{u}_{L\,i}$--$d_{R\,j}$--$H^+ $,
$\bar{u}_{R\,i}$--$d_{L\,j}$--$H^+$ ($i,j=1,2,3$)
are weighted by the elements of the CKM matrix $V=V_u^\dagger V_d$. 
The supersymmetric counterpart of these vertices, 
$ \bar{u}_{L\,i}$--${\widetilde{D}}_{j}$--${\widetilde W}^+$,
$ \bar{u}_{L\,i}$--${\widetilde{D}}_{j}^{c\,\ast}$--$ {\widetilde H}^+$, 
$ \bar{u}_{R\,i}$--${\widetilde{D}}_{j}$--${\widetilde H}^-$, 
are also proportional to $V_{ij}$ in the limit of unbroken 
supersymmetry.

To illustrate the sources of flavour violation that  may be present in
supersymmetric models in addition to those encoded in the CKM matrix,
it is instructive to consider in detail the contributions to the 
squared-mass matrix of a squark of flavour $f$.  
The relation between off-diagonal terms in this squared-mass matrix 
and the type of operators inducing the decay $b \to s \gamma$, 
will then become clear. 
Since present collider limits give indications that 
the squark masses are larger than those of the corresponding 
quarks, the largest entries in the squark mass matrices squared
must come from the soft potential, directly 
linked to the mechanism of supersymmetry breaking. 
When restricted to the terms relevant to squark masses
and quark-flavour transitions, the soft potential can be 
expressed in terms of the current eigenstates scalar fields as:
\begin{eqnarray}
\lefteqn{
 V_{soft} \quad \supset \quad  
                                {\widetilde{Q}}_i^{o\,\ast}     \,
   m_{\,{\widetilde{Q}}\,ij}^{\,2}{\widetilde{Q}}_j^o           \, + \,
                                {\widetilde{D}}_i^{c\,o\,\ast}  \,
   m_{\,{\widetilde{D}}\,ij}^{\,2}{\widetilde{D}}_j^{c\,o}      \, + \,
                                {\widetilde{U}}_i^{c\,o\,\ast}  \,
   m_{\,{\widetilde{U}}\,ij}^{\,2}{\widetilde{U}}_j^{c\,o}      \  + \ 
\left(  -  \frac{1}{2} M_3 \lambda_3 \lambda_3                  \  + \right.   
    } \nonumber \\ &&  \hspace*{6truecm} 
  A_{d,ij} H_d \,{\widetilde{Q}}_i^o {\widetilde{D}}_j^{c\,o}   \, + \, 
  A_{u,ij} H_u \,{\widetilde{Q}}_i^o {\widetilde{U}}_j^{c\,o}   \, +
  \ {\rm h.c.} \Biggr)\,.
\label{vsoft}
\end{eqnarray}
In~(\ref{vsoft}), $m_{\,{\widetilde{Q}}}^2$, $m_{\,{\widetilde{D}}}^2$, 
and $m_{\,{\widetilde{U}}}^2$ are hermitian matrices. The gluino
$\tilde{g}$, a four-component Majorana spinor, is expressed in terms of 
the Weyl spinor $\lambda_3$ and has mass $m_{\tilde{g}} = M_3$. 
Notice that, for the trilinear terms
$ A_{d,ij} H_d {\widetilde Q}_i^o {\widetilde D}_j^{c\,o}$, 
no proportionality to the Yukawa couplings is assumed. These trilinear 
scalar terms are left
completely general and may also represent non-holomorphic ones, 
of the type
$A^\prime _{d,ij} H_u^\ast {\widetilde Q}_i^o {\widetilde D}_j^{c\,o}$ 
discussed in~\cite{BFPT}.

Thus, in the interaction basis
$({\widetilde Q}^{o}_1,  {\widetilde Q}^{o}_2, {\widetilde Q}^{o}_3,
  {\widetilde Q}^{c\,o\,\ast}_1, {\widetilde Q}^{c\,o\,\ast}_2,
  {\widetilde Q}^{c\,o\,\ast}_3)$, 
often denoted also as 
$({\widetilde q}^o_{L\,1},$ ${\widetilde q}^o_{L\,2},$ 
${\widetilde q}^o_{L\,3},$ ${\widetilde q}^o_{R\,1},$ 
${\widetilde q}^o_{R\,2},$ 
${\widetilde q}^o_{R\,3})$, 
the squared-mass matrix for a squark of flavour $f$ has the form 
\begin{equation}
{\cal M}_f^2 \equiv  \left( \begin{array}{cc}
  m^2_{\,f,\,LL} +F_{f\,LL} +D_{f\,LL}           & 
                 \left(m_{\,f,\,LR}^2\right) + F_{f\,LR} 
                                                     \\[1.01ex]
 \left(m_{\,f,\,LR}^{2}\right)^{\dagger} + F_{f\,RL} &
             \ \ m^2_{\,f,\,RR} + F_{f\,RR} +D_{f\,RR}                
 \end{array} \right) \,.
\label{squarkmassmatr}
\end{equation}
The term $ m^2_{\,f,\,LL}$ is $m^2_{\,\widetilde{Q}}\,$,
for both, up- and down-type squarks; 
$ m^2_{\,f,\,RR}$ is 
$ m^2_{\,\widetilde{D}}$ for a down-type squark and 
$ m^2_{\,\widetilde{U}}$  for an up-type squark.
The off-diagonal $3 \times 3$ block matrix $ m_{\,f,\,LR}^2$ is 
$A^{\,\ast}_{d} v_d $ for a down squark, 
$A^{\,\ast}_{u} v_u $ for an up-type one. 
(The two vacuum expectation values are chosen to be real.)
It should be stressed that, differently from $ m^2_{\,f,\,LL}$
and $ m^2_{\,f,\,RR}$, the off-diagonal $3 \times 3$ matrix 
$m_{\,f,\,LR}^2$ is not hermitian. In other words, it is 
$A_{d, ij} \ne A^{\,\ast}_{d,ji}$ as well as 
$A_{u, ij} \ne A^{\,\ast}_{u,ji}$.

The $D$-term contributions $D_{f\,LL}$ and $D_{f\,RR}$ 
to the squared-mass matrix~(\ref{squarkmassmatr}), 
\begin{equation}
 D_{f\,LL,RR} =  \cos 2\beta \, M_Z^2 
   \left(T_f^3 - Q_f \sin^2\theta_W \right) 
{{\mathchoice {\rm 1\mskip-4mu l} {\rm 1\mskip-4mu l}
{\rm 1\mskip-4.5mu l} {\rm 1\mskip-5mu l}}}_3\,,
\label{dterm}
\end{equation}
are diagonal in flavour space.

The explicit form for the $F$-term contributions 
can be obtained from scalar quartic
couplings arising from the superpotential~(\ref{superpot}):
\begin{eqnarray}
V_{F}                                                 & \ \supset \ &  
 v_d^2\, {\widetilde{D}}^{o\,\ast}_i    \left(h_d^\dagger h_d \right)_{ij}
         {\widetilde{D}}^{o}_j                        \ + \
 v_d^2\, {\widetilde{D}}^{c\,o}_i  \left(h_d h_d^\dagger \right)_{ij}
         {\widetilde{D}}^{c\,o\,\ast}_j               \ - \
 \left( \mu \, v_u \, 
         {\widetilde{D}}^{o\,\ast}_i     h_{d,ij}^\dagger
         {\widetilde{D}}^{c\,o\,\ast}_j 
 \, + \ {\rm h.c.}\, \right)      +                   \nonumber \\
                                                      &  &   
 v_u^2\, {\widetilde{U}}^{o\,\ast}_i    \left(h_u^\dagger h_u \right)_{ij}
         {\widetilde{U}}^{o}_j                        \ + \
 v_u^2\, {\widetilde{U}}^{c\,o}_i \left(h_u  h_u^\dagger \right)_{ij}
         {\widetilde{U}}^{c\,o\,\ast}_j               \ - \
 \left( \mu \, v_d \, 
         {\widetilde{U}}^{o\,\ast}_i  h_{u,ij}^\dagger
         {\widetilde{U}}^{c\,o\,\ast}_j 
 \, + \ {\rm h.c.}\, \right)\,.
\label{fterm}
\end{eqnarray}
The rotation~(\ref{superotation}) reduces $F_{f\,LL}$ and 
$F_{f\,RR}$ to their diagonal form
$$
 m^2_{d\,i} {\widetilde{D}}^{\ast}_i {\widetilde{D}}_i\,,  
\hspace*{0.5truecm}
 m^2_{u\,i} {\widetilde{U}}^{\ast}_i {\widetilde{U}}_i\,, 
\hspace*{0.5truecm}
 m^2_{d\,i} {\widetilde{D}}^{c\,\ast}_i {\widetilde{D}}^{c}_i\,,  
\hspace*{0.5truecm}
 m^2_{u\,i} {\widetilde{U}}^{c\,\ast}_i {\widetilde{U}}^{c}_i\,, 
$$ 
as well as $F_{f\,LR}$ ($F_{f\,RL}=F_{f\,LR}^\dagger$) to 
$$  -\mu (m_{d,i} \tan \beta)
  {\widetilde{D}}^{\ast}_i  {\widetilde{D}}^{c\,\ast}_i ,
\hspace*{0.5truecm}
  -\mu (m_{u,i} \cot \beta)
  {\widetilde{U}}^{\ast}_i  {\widetilde{U}}^{c\,\ast}_i. 
$$

Therefore, once up- and down-quarks are brought to their mass 
eigenstate basis through the rotation~(\ref{superotation}), the 
only sources of flavour violation in the squark sector arise from
the off-diagonal terms in the soft mass matrices  
$ m^2_{\,f,\,LL}$, $ m^2_{\,f,\,RR}$, and 
$ m^2_{\,f,\,LR}$~\footnote{No new symbols are introduced to indicate 
 the unknown matrices 
 $ m^2_{\,f,\,LL}$, $ m^2_{\,f,\,RR}$, and 
 $ m^2_{\,f,\,LR}$ after the rotation~(\ref{superotation}). Notice, 
 however, that $m^2_{\,u,\,LL}$ and $m^2_{\,d,\,LL}$, equal before 
 this rotation, are now related as  
 $m^2_{\,u,\,LL} = V m^2_{d,LL} V^\dagger$.}
Their origin, as their 
magnitude, is a model-dependent matter based on the interplay
between the dynamics of flavour and that dictating the breaking of
supersymmetry.  In general, however, they give rise to large flavour--quark
transitions at the loop level, through large couplings of gluinos 
to quarks and squarks belonging to different generations.

One very drastic approach to this supersymmetric flavour problem is
that of mSUGRA. In this model (or class of models) the soft
potential~(\ref{vsoft}) is characterized at some high scale, typically
a grand unification scale, by the universality of the scalar masses:
\begin{equation}
 m_{\,{\widetilde{Q}}\,ij} \,=\, 
 m_{\,{\widetilde{U}}\,ij} \,=\,  m_{\,{\widetilde{D}}\,ij} \,=\, 
 \widetilde{m} \,\delta_{ij}\,;  
\label{universality}
\end{equation}
and the proportionality of the trilinear terms to the Yukawa couplings, 
through a universal parameter $A$:
\begin{equation}
  A_{d,ij} = A h_{d,ij}\,;    \hspace*{1.0truecm}
  A_{u,ij} = A h_{u,ij}.
\label{proportionality}    
\end{equation}
At this high scale, the only source of flavour violation is contained 
in the superpotential, indicating that the breaking of supersymmetry
occurs at a scale where the dynamics of flavour has already taken
place.

An elegant solution to the flavour problem is obtained in GMSB models,
in which the signal of supersymmetry breaking is transmitted to the
visible sector of fields ${\widetilde{Q}}^o$, ${\widetilde{U}}^o$,
${\widetilde{D}}^o$, $H_1, H_2$, etc.,  by flavour-blind gauge
interactions. In these models, at the scale of supersymmetry breaking,
all matrices in~(\ref{universality}) are diagonal, although different,
and the common value of $A$ in~(\ref{proportionality}) is set to zero.

In both mSUGRA and GMSB models, sources of flavour violation in the
scalar sector are generated radiatively at the electroweak scale
through the scalar quartic couplings proportional to Yukawa matrices.
A simple inspection shows that intergenerational mixing terms due to 
only one type of Yukawa matrix, get eliminated by the
rotation~(\ref{superotation}): no off-diagonal terms are therefore
possible in $ m^2_{\,f,\,RR}$ in these models. On the contrary,
flavour-violating terms are not rotated away in the $ m^2_{\,f,\,LL}$
sector in which radiative contributions arise from quartic scalar
couplings proportional to both matrices $h_d$ and $h_u$. Being
loop-induced, this source of flavour violation is, in general,
small~\cite{BBMR}, but it becomes non-negligible for large values of
$\tan \beta$~\cite{FB}. By this reasoning it becomes clear that, while
a contribution to the operator ${\cal O}_{7b,\tilde{g}}$ can arise
from an off-diagonal term mixing the second- and third-generation left
squarks $(m^2_{\,d,\,LL})_{23}$, as shown in Fig.~\ref{glopb}, no
contribution to ${\cal O}_{7b,\tilde{g}}^\prime$ is possible in mSUGRA
and GMSB models.  The same holds for all other primed operators. These
operators may nevertheless acquire non-vanishing contributions in more 
general models, in which, for example, there exists an off-diagonal term
$(m^2_{\,d,\,RR})_{23}$.

Also vanishing, in mSUGRA and GMSB models, is the contribution to the
operator ${\cal O}_{7\tilde{g},\tilde{g}}$ coming from a left--right
mixing element $(m^2_{\,d,\,LR})_{23}$. A contribution to this
operator can however be induced, even in these models, by 
intergenerational mixing terms in $ m^2_{\,d,\,LL}$,
$(m^2_{\,d,\,LL})_{23}$, and the flavour-diagonal left--right term
$(m^2_{\,d,\,LR})_{33}$.  In the mass-insertion formalism, 
often used for the calculation of supersymmetric contributions to 
FCNC processes~\cite{DGH}, the first non-vanishing contribution to 
${\cal O}_{7\tilde{g},\tilde{g}}$ is then generated by the double 
insertion shown in the first diagram of Fig.~\ref{glopgtwoins}.
It will be shown later that, in generic supersymmetric models, 
this contribution to ${\cal O}_{7\tilde{g},\tilde{g}}$ turns out to 
give the strongest constraint on $(m^2_{\,d,\,LL})_{23}$, when 
reasonable values of $(m^2_{\,d,\,LR})_{33}$ are chosen.

As advocated in the introduction, the aim of this paper is to 
provide a calculation as general as possible of the 
gluino contribution to the decay $b \to s \gamma$, i.e. a 
calculation that  applies to supersymmetric models with the most 
general soft terms. The QCD-corrected branching ratio for this
decay can then be used to constrain the size of the off-diagonal 
elements of the mass matrices $ m^2_{\,d,\,LL}$, $ m^2_{\,d,\,RR}$, 
and $m^2_{\,d,\,LR}$. Since different operators contribute to this
decay, with different numerical impact on its rate, some of these
flavour-violating terms may turn out to be poorly constrained. Thus, 
given the generality of such a calculation, it is 
convenient to rely on the mass eigenstate 
formalism, which remains valid even when the intergenerational mixing 
elements are large. The procedure used follows closely 
Refs.~\cite{GGRZ,BBMR}. The diagonalization of the two $6 \times 6$ 
squark mass matrices squared ${\cal M}^2_d$ and ${\cal M}^2_u$  
yields the eigenvalues $m_{\tilde{d}_k}^2$ and $m_{\tilde{u}_k}^2$ 
($k=1,...,6$). The corresponding mass eigenstates, 
$\tilde{u}_{k}$ and $\tilde{d}_{k}$ ($k=1,...,6$) are 
related to the fields $\tilde{u}_{Lj}$, $\tilde{u}_{Rj}$
and $\tilde{d}_{Lj}$, $\tilde{d}_{Rj}$
($j=1,...,3$) as: 
\begin{equation}
\tilde{u}_{L,R} = \Gamma^\dagger_{UL,R} \, \tilde{u} \,,
\hspace*{1truecm}
\tilde{d}_{L,R} = \Gamma^\dagger_{DL,R} \, \tilde{d} \,,
\label{qdiag}
\end{equation}
where the four matrices
$\Gamma_{UL,R}$ and $\Gamma_{DL,R}$ are $6 \times 3$ mixing
matrices. The gluino--quark--squark vertices are 
explicitly given in Ref.~\cite{BBMR}.

\section{Wilson Coefficients at the Electroweak Scale}
\label{WilsonCoeff}

At the matching scale $\mu_W$, the non-vanishing Wilson coefficients
for the SM operators in eqs.~(\ref{smfourquarkop}) and~(\ref{smmagnop})
are, at leading order in $\alpha_s$:
\begin{eqnarray}
 C_{2}(\mu_W)  & = & \ 1                  
\nonumber \\[1.5ex]
 C_{7}(\mu_W)  & = & \frac{x_{tw}}{24\,(x_{tw}-1)^4} \, \left(
 {-8x_{tw}^3+3x_{tw}^2+12x_{tw}-7+(18x_{tw}^2-12x_{tw}) \ln x_{tw}}
                                                        \right) 
\nonumber \\[1.5ex]    
 C_{8}(\mu_W)  & = & \frac{x_{tw}}{\ 8\,(x_{tw}-1)^4} \, \left(
 {-x_{tw}^3+6x_{tw}^2-3x_{tw}-2-6x_{tw} \ln x_{tw}}      \right) 
\,,
\label{wclosm}                                          
\end{eqnarray}
with $x_{tw} \equiv m_t^2/M_W^2$.

Among the coefficients arising from the virtual exchange of a
gluino at the matching scale, the non-vanishing ones 
are~\footnote{The linear combination 
 $C_{7b,\tilde{g}}(\mu_W) {\cal O}_{7b,\tilde{g}}(\mu_W) +
 C_{7\tilde{g},\tilde{g}}(\mu_W) {\cal O}_{7\tilde{g},\tilde{g}}(\mu_W)$
 coincides with the expression $\delta C_7(\mu_W) {\cal O}_7(\mu_W)$
 given in the  literature (see e.g.~\cite{BBMR,CHO}), where ${\cal O}_7$
 is the standard model operator.}:
\begin{eqnarray}
C_{7b,\tilde{g}}(\mu_W)              & = &
\ \ 
-\frac{e_d}{16 \pi^2} \ C(R)
 \sum_{k=1} ^6 \frac{1}{m_{\tilde{d}_k}^2} 
\left( \Gamma_{DL}^{kb} \, \Gamma_{DL}^{\ast\,ks} \right)
 F_2(x_{gd_k}) 
                                     \nonumber \\
C_{7\tilde{g},\tilde{g}}(\mu_W)      & = & 
 m_{\tilde g}\,
 \frac{e_d}{16 \pi^2} \ C(R) 
 \sum_{k=1} ^6 \frac{1}{m_{\tilde{d}_k}^2} 
\left( \Gamma_{DR}^{kb} \, \Gamma_{DL}^{\ast\,ks} \right)
 F_4(x_{gd_k})\,, 
\label{phgl}                
\end{eqnarray}
in the case of magnetic operators and 
\begin{eqnarray}
C_{8b,\tilde{g}}(\mu_W)              & = &
\ \ 
-\frac{1}{16 \pi^2} 
 \sum_{k=1} ^6 \frac{1}{m_{\tilde{d}_k}^2} 
\left( \Gamma_{DL}^{kb} \, \Gamma_{DL}^{\ast\,ks} \right) \,  
\left[
\left(C(R) -\!{\scriptstyle{1\over 2}} C(G)\right) F_2(x_{gd_k})
      -{\scriptstyle{1\over 2}} C(G) F_1(x_{gd_k}) 
\right]
                                     \nonumber \\
C_{8\tilde{g},\tilde{g}}(\mu_W)      & = & 
 m_{\tilde g}\, 
 \frac{1}{16 \pi^2} 
 \sum_{k=1} ^6 \frac{1}{m_{\tilde{d}_k}^2} 
\left( \Gamma_{DR}^{kb} \, \Gamma_{DL}^{\ast\,ks} \right) \,            
\left[
\left(C(R) -\!{\scriptstyle{1\over 2}} C(G)\right) F_4(x_{gd_k})
      -{\scriptstyle{1\over 2}} C(G) F_3(x_{gd_k}) 
\right] \,,
\label{glgl}                
\end{eqnarray}
in the case of chromomagnetic operators. The coefficients
$C_{7\tilde{g},\tilde{g}}(\mu_W)$ and
$C_{8\tilde{g},\tilde{g}}(\mu_W)$ of higher dimensionality to
compensate the lower dimensionality of the corresponding operators.
The ratios $x_{gd_k}$ are now defined as 
$x_{gd_k} \equiv m_{\tilde g}^2/m_{\tilde{d}_k}^2$; the Casimir
factors $C(R)$ and $C(G)$ are respectively $C(R)=4/3$ and $C(G)= 3$;
and the functions $F_i(x)$, $i = 1,...,4$, are given in
Appendix~\ref{functions}.  The Wilson coefficients of the
corresponding primed operators are obtained through the interchange
$\Gamma_{DR}^{ij} \leftrightarrow \Gamma_{DL}^{ij} $ in
eqs.~(\ref{phgl}) and~(\ref{glgl}).  The coefficients of the magnetic
and chromomagnetic operators, proportional to the $c$-quark mass, vanish
at the matching scale at lowest order in $\alpha_s$.

Compared to the SM,
there is a larger number of magnetic and chromomagnetic operators with
different chirality and dimensionality. The different chiralities are
due to the fact that the gluino couples
both to left- and right handed quarks and the associated squarks.
In contrast, the $W$ has only left-handed couplings and therefore right
handed fields only arise if their masses are not neglected; usually
only (chromo)magnetic operators with right-handed $b$-quarks are included. 
Similarly, the occurence
of (chromo)magnetic operators with differing dimensions can also be understood
from the chirality structure of the gluino couplings.
Some of the new operators differ from the SM (chromo)magnetic operators
only by an additional factor $g_s^2$. These were introduced as additional
operators for practical reasons.

Penguin diagrams mediated by the virtual exchange of a gluino and a
gluon, yield non-vanishing coefficients only for the operators 
${\cal O}_{11,\tilde{g}}^q$--${\cal O}_{14,\tilde{g}}^q$:
\begin{eqnarray}
 C_{11,\tilde{g}}^q(\mu_W)           & = &  
\, \frac{1}{16 \pi^2}\, \frac{1}{3} \, 
 \sum_{k=1} ^6 \frac{1}{m_{\tilde{d}_k}^2} 
\left( \Gamma_{DL}^{kb} \, \Gamma_{DL}^{\ast\,ks} \right) \,            
\left[
\left(C(R) -\!{\scriptstyle{1\over 2}} C(G)\right) F_6(x_{gd_k})
      +{\scriptstyle{1\over 2}} C(G) F_5(x_{gd_k}) 
\right]
                                     \nonumber \\
 C_{12,\tilde{g}}^q(\mu_W)           & = &   
 - \frac{1}{16 \pi^2} \ 
 \sum_{k=1} ^6 \frac{1}{m_{\tilde{d}_k}^2} 
\left( \Gamma_{DL}^{kb} \, \Gamma_{DL}^{\ast\,ks} \right) \,            
\left[
\left(C(R) -\!{\scriptstyle{1\over 2}} C(G)\right) F_6(x_{gd_k})
      +{\scriptstyle{1\over 2}} C(G) F_5(x_{gd_k}) 
\right]
                                     \nonumber \\[1.01ex]
 C_{13,\tilde{g}}^q(\mu_W)           & = & 
          C_{11,\tilde{g}}^q(\mu_W)
                                     \nonumber \\[1.1ex]
 C_{14,\tilde{g}}^q(\mu_W)           & = & 
          C_{12,\tilde{g}}^q(\mu_W) \,, 
\label{pengcoeff}
\end{eqnarray}
as well as coefficients for the corresponding primed operators,
${\cal O}_{11,\tilde{g}}^{q\,\prime}$--${\cal O}_{14,\tilde{g}}^{q\,\prime}$,
which can be obtained from those in eq.~(\ref{pengcoeff}) 
by interchanging $\Gamma_{DR}^{ij} \leftrightarrow \Gamma_{DL}^{ij}$. 
These coefficients are actually independent of the quark label $q$.

Box diagrams~\footnote{Note that these diagrams are finite and all 
 the manipulations needed to eliminate the charge conjugation
 matrices in the crossed topologies shown in Figs.~(\ref{boxdiagr})
 are well defined.}, with exchange of two virtual gluinos, yield 
the following contributions to the coefficients 
$C_{11,\tilde{g}}^q$--$C_{14,\tilde{g}}^q$:
\begin{eqnarray}
C_{11,\tilde{g}}^q(\mu_W)            & = &
\frac{1}{16 \pi^2}\,\frac{1}{m_{\tilde{g}}^2} \sum_{k,h=1} ^6  
\left\{\frac{1}{36} \, 
\left(\Gamma_{D\,L}^{\,kb}    \,\Gamma_{D\,L}^{\ast\,ks} \right)
\left(\Gamma_{Q\,L}^{\ast\,hq}\,\Gamma_{Q\,L}^{\,hq}     \right)
\left[G(x_{d_kg},x_{q_hg})-20F(x_{d_kg},x_{q_hg})\right]
\right. \nonumber \\ & & \left.
\phantom{\frac{1}{16 \pi^2}\,\frac{1}{m_{\tilde{g}}^2} }
+\,\delta_{qd} \,
 \frac{1}{12} \,           
\left(\Gamma_{D\,L}^{\,kb}    \,\Gamma_{D\,L}^{\ast\,hs} \right)
\left(\Gamma_{Q\,L}^{\ast\,kq}\,\Gamma_{Q\,L}^{\,hq}     \right)
\left[7G(x_{d_kg},x_{q_hg}) +4 F(x_{d_kg},x_{q_hg})\right] 
\right\}
                                     \nonumber \\[1.01ex]
C_{12,\tilde{g}}^q(\mu_W)            & = & 
\frac{1}{16 \pi^2}\,\frac{1}{m_{\tilde{g}}^2} \sum_{k,h=1} ^6  
\left\{\frac{1}{12} \, 
\left(\Gamma_{D\,L}^{kb}       \,\Gamma_{D\,L}^{\ast\,ks} \right)
\left(\Gamma_{Q\,L}^{\ast\,hq} \,\Gamma_{Q\,L}^{hq}       \right)
\left[7 G(x_{d_kg},x_{q_hg})+4 F(x_{d_kg},x_{q_hg})\right]    
\right. \nonumber \\ & & \left.
\phantom{\frac{1}{16 \pi^2}\,\frac{1}{m_{\tilde{g}}^2}}
+\,\delta_{qd} \,
\frac{1}{36}\,
\left(\Gamma_{D\,L}^{kb}       \,\Gamma_{D\,L}^{\ast\,hs} \right)
\left(\Gamma_{Q\,L}^{\ast\,kq} \,\Gamma_{Q\,L}^{hq}       \right)
\left[G(x_{d_kg},x_{q_hg})-20 F(x_{d_kg},x_{q_hg})\right] 
\right\}   
                                     \nonumber \\[1.01ex]
C_{13,\tilde{g}}^q(\mu_W)            & = & 
\frac{1}{16 \pi^2}\,\frac{1}{m_{\tilde{g}}^2} \sum_{k,h=1} ^6  
\left\{-\frac{1}{18} \, 
\left(\Gamma_{D\,L}^{kb}       \,\Gamma_{D\,L}^{\ast\,ks} \right)
\left(\Gamma_{Q\,R}^{\ast\,hq} \,\Gamma_{Q\,R}^{hq}       \right)
\left[5 G(x_{d_kg},x_{q_hg})-F(x_{d_kg},x_{q_hg})\right]
\right. \nonumber \\ & & \left.
\phantom{\frac{1}{16 \pi^2}\,\frac{1}{m_{\tilde{g}}^2} -}
+\,\delta_{qd} \,
\frac{5}{12}\, 
\left(\Gamma_{D\,L}^{kb}       \,\Gamma_{D\,L}^{\ast\,hs} \right)
\left(\Gamma_{Q\,R}^{\ast\,kq} \,\Gamma_{Q\,R}^{hq}       \right)
G(x_{d_kg},x_{q_hg})
\right\}   
                                     \nonumber \\[1.01ex]
C_{14,\tilde{g}}^q(\mu_W)            & = &
\frac{1}{16 \pi^2}\,\frac{1}{m_{\tilde{g}}^2} \sum_{k,h=1} ^6  
\left\{\ \frac{1}{6} \, 
\left(\Gamma_{D\,L}^{kb}      \,\Gamma_{D\,L}^{\ast\,ks} \right)
\left(\Gamma_{Q\,R}^{\ast\,hq}\,\Gamma_{Q\,R}^{hq}       \right)
\left[G(x_{d_kg},x_{q_hg})+7F(x_{d_kg},x_{q_hg})\right]
\right. \nonumber \\ & & \left.
\phantom{\frac{1}{16 \pi^2}\,\frac{1}{m_{\tilde{g}}^2}}
+\,\delta_{qd} \,
\frac{11}{36}\, 
\left(\Gamma_{D\,L}^{kb}      \,\Gamma_{D\,L}^{\ast\,hs} \right)
\left(\Gamma_{Q\,R}^{\ast\,kq}\,\Gamma_{Q\,R}^{hq}       \right)
G(x_{d_kg},x_{q_hg})
\right\} \,,  
\label{wcoeffbox14}
\end{eqnarray}
with the corresponding primed coefficients obtained through the
interchange $\Gamma_{DL}^{ij} \leftrightarrow \Gamma_{DR}^{ij} $ 
and $\Gamma_{QL}^{ij} \leftrightarrow \Gamma_{QR}^{ij}$.
Notice that the symbol $\delta_{qd}$ is the Kronecker delta, equal to
one when $q$ is the down-quark and zero when $q$ is a different 
quark.
For $q=d$, also, the subscript $Q$ in the two 
combinations 
$(\Gamma_{Q\,L}^{\ast\,hq}\,\Gamma_{Q\,L}^{hq})$ and
$(\Gamma_{Q\,R}^{\ast\,hq}\,\Gamma_{Q\,R}^{hq})$ has to be identified with
$D$, typical of a down-type squark exchanged in the box diagram. 
The box-diagram 
functions $G(x,y)$ and $F(x,y)$ are explicitly listed in 
Appendix~\ref{functions}.

The remaining coefficients $C_{15,\tilde{g}}^q$--$C_{20,\tilde{g}}^q$,
in mass insertion language, are characterized by an odd number of
$L$--$R$ insertions in each squark line. In the mass-eigenstate basis
used for squarks in this analysis, they are:
\begin{eqnarray}
C_{15,\tilde{g}}^q(\mu_W)            & = &
\frac{1}{16 \pi^2}\,\frac{1}{m_{\tilde{g}}^2} \sum_{k,h=1} ^6  
\left\{\frac{11}{18} \, 
\left(\Gamma_{D\,R}^{kb}      \,\Gamma_{D\,L}^{\ast\,ks} \right)
\left(\Gamma_{Q\,L}^{\ast\,hq}\,\Gamma_{Q\,R}^{hq}       \right)
\,       F(x_{d_kg},x_{q_hg})
\right. \nonumber \\ & & \left.
\phantom{\frac{1}{16 \pi^2}\,\frac{1}{m_{\tilde{g}}^2}}
\ -\,\delta_{qd} \,
\frac{8}{3}\, 
\left(\Gamma_{D\,R}^{kb}      \,\Gamma_{D\,L}^{\ast\,hs} \right)
\left(\Gamma_{Q\,L}^{\ast\,kq}\,\Gamma_{Q\,R}^{hq}       \right)
F(x_{d_kg},x_{q_hg})
\right\}   
                                     \nonumber \\[1.01ex]
C_{16,\tilde{g}}^q(\mu_W)            & = &
\frac{1}{16 \pi^2}\,\frac{1}{m_{\tilde{g}}^2} \sum_{k,h=1} ^6  
\left\{\frac{5}{6} \, 
\left(\Gamma_{D\,R}^{kb}      \,\Gamma_{D\,L}^{\ast\,ks}  \right)
\left(\Gamma_{Q\,L}^{\ast\,hq}\,\Gamma_{Q\,R}^{hq}        \right)
\,       F(x_{d_kg},x_{q_hg})
\right. \nonumber \\ & & \left.
\phantom{\frac{1}{16 \pi^2}\,\frac{1}{m_{\tilde{g}}^2}}
+\,\delta_{qd} \,
\frac{4}{9} \, 
\left(\Gamma_{D\,R}^{kb}      \,\Gamma_{D\,L}^{\ast\,hs}  \right)
\left(\Gamma_{Q\,L}^{\ast\,kq}\,\Gamma_{Q\,R}^{hq}        \right)
F(x_{d_kg},x_{q_hg})
\right\}   
                                     \nonumber \\[1.01ex]
C_{17,\tilde{g}}^q(\mu_W)            & = & 
\frac{1}{16 \pi^2}\,\frac{1}{m_{\tilde{g}}^2} \sum_{k,h=1} ^6  
\left\{-\frac{11}{18} \, 
\left(\Gamma_{D\,R}^{kb}      \,\Gamma_{D\,L}^{\ast\,ks} \right)
\left(\Gamma_{Q\,R}^{\ast\,hq}\,\Gamma_{Q\,L}^{hq}       \right)
\,   G(x_{d_kg},x_{q_hg}) 
\right. \nonumber \\ & & \left.
\phantom{\frac{1}{16 \pi^2}\,\frac{1}{m_{\tilde{g}}^2} -}
\ -\,\delta_{qd} \, 
\frac{1}{3} \,
\left(\Gamma_{D\,R}^{kb}      \,\Gamma_{D\,L}^{\ast\,hs} \right)
\left(\Gamma_{Q\,R}^{\ast\,kq}\,\Gamma_{Q\,L}^{hq}       \right)
\left[G(x_{d_kg},x_{q_hg}) + 7 F(x_{d_kg},x_{q_hg})\right]
\right\}        
                                     \nonumber \\[1.01ex]
C_{18,\tilde{g}}^q(\mu_W)            & = &
\frac{1}{16 \pi^2}\,\frac{1}{m_{\tilde{g}}^2} \sum_{k,h=1} ^6  
\left\{-\frac{5}{6} \, 
\left(\Gamma_{D\,R}^{kb}       \,\Gamma_{D\,L}^{\ast\,ks} \right)
\left(\Gamma_{Q\,R}^{\ast\,hq} \,\Gamma_{Q\,L}^{hq}       \right)
 \,     G(x_{d_kg},x_{q_hg})
\right. \nonumber \\ & & \left.
\phantom{\frac{1}{16 \pi^2}\,\frac{1}{m_{\tilde{g}}^2} -}
+\,\delta_{qd} \,
\frac{1}{9}
\left(\Gamma_{D\,R}^{kb}       \,\Gamma_{D\,L}^{\ast\,hs} \right)
\left(\Gamma_{Q\,R}^{\ast\,kq} \,\Gamma_{Q\,L}^{hq}       \right)
\left[5 G(x_{d_kg},x_{q_hg}) - F(x_{d_kg},x_{q_hg})\right]
\right\}        
                                     \nonumber \\[1.01ex]
C_{19,\tilde{g}}^q(\mu_W)            & = & 
\frac{1}{16 \pi^2}\,\frac{1}{m_{\tilde{g}}^2} \sum_{k,h=1} ^6  
\left\{-\frac{1}{8} \, 
\left(\Gamma_{D\,R}^{kb}       \,\Gamma_{D\,L}^{\ast\,ks} \right)
\left(\Gamma_{Q\,L}^{\ast\,hq} \,\Gamma_{Q\,R}^{hq}       \right)
\,       F(x_{d_kg},x_{q_hg})
\right. \nonumber \\ & & \left.
\phantom{\frac{1}{16 \pi^2}\,\frac{1}{m_{\tilde{g}}^2} -}
+\,\delta_{qd} \,
\frac{1}{12} 
\left(\Gamma_{D\,R}^{kb}       \,\Gamma_{D\,L}^{\ast\,hs} \right)
\left(\Gamma_{Q\,L}^{\ast\,kq} \,\Gamma_{Q\,R}^{hq}       \right)
F(x_{d_kg},x_{q_hg})
\right\}   
                                     \nonumber \\[1.01ex]
C_{20,\tilde{g}}^q(\mu_W)            & = & 
\frac{1}{16 \pi^2}\,\frac{1}{m_{\tilde{g}}^2} \sum_{k,h=1} ^6  
\left\{\ \frac{3}{8} \, 
\left(\Gamma_{D\,R}^{kb}       \,\Gamma_{D\,L}^{\ast\,ks}  \right)
\left(\Gamma_{Q\,L}^{\ast\,hq} \,\Gamma_{Q\,R}^{hq}        \right)
\,      F(x_{d_kg},x_{q_hg})  
\right. \nonumber \\ & & \left.
\phantom{\frac{1}{16 \pi^2}\,\frac{1}{m_{\tilde{g}}^2}}
-\,\delta_{qd} \,
\frac{5}{36} \, 
\left(\Gamma_{D\,R}^{kb}       \,\Gamma_{D\,L}^{\ast\,hs}  \right)
\left(\Gamma_{Q\,L}^{\ast\,kq} \,\Gamma_{Q\,R}^{hq}        \right)
F(x_{d_kg},x_{q_hg})
\right\}\,.   
\label{wcoeffbox}
\end{eqnarray}
The considerations made for the coefficients~(\ref{wcoeffbox14})
hold also here: 
the corresponding primed coefficients are obtained through the
interchanges $\Gamma_{DL}^{ij} \leftrightarrow \Gamma_{DR}^{ij} $ 
and $\Gamma_{QL}^{ij} \leftrightarrow \Gamma_{QR}^{ij}$, and 
$\delta_{qd}$ always vanishes, except for $q=d$. 
Under renormalization, the operators corresponding to the 
coefficients~(\ref{wcoeffbox}) mix with the magnetic and chromomagnetic 
operators in~(\ref{gmagnopb}) and~(\ref{gmagnopc}) by undergoing a 
chirality flip proportional to $m_q$. Therefore, only $q=b$ and $q=c$
can contribute to the decay $b \to s \gamma$ in the approximation
of massless light quarks made here.

\section{Wilson Coefficients at the Decay Scale}
\label{gmatrix}

As already mentioned in Sec.~\ref{ordersplit}, the two terms 
${\cal H}_{eff}^{W}$ and ${\cal H}_{eff}^{\tilde{g}}$ in the effective
Hamiltonian (\ref{hfull}) undergo separate renormalization. The
anomalous-dimension matrix of the SM operators 
${\cal O}_1$--${\cal O}_8$ and the evolution of the corresponding
Wilson coefficients to the decay scale $\mu_b$ are very well known and
can be found in~\cite{BG}.

The evolution of the gluino-induced Wilson coefficients 
$C_{i,\tilde{g}}$ from the matching scale $\mu_W$  down to the 
low-energy scale $\mu_b$ is described by the  
renormalization group equation:
\begin{equation}
\label{RGE}
\mu \frac{d}{d\mu} C_{i,\tilde{g}} =  C_{j,\tilde{g}}(\mu) \, 
 \gamma_{ji,\tilde{g}}(\mu) \,.
\end{equation}
The usual perturbative expansion for the initial conditions of 
the Wilson coefficients, 
\begin{equation}
 C_{i,\tilde{g}} (\mu_W) =       C_{i,\tilde{g}}^{0}(\mu_W) + 
 \frac{\alpha_s(\mu_W)}{4\pi} \, C_{i,\tilde{g}}^{1}(\mu_W)  + .....
\label{coeffdecomp}
 \,,
\end{equation}
as well as for the elements of $ \gamma_{ji\,\tilde{g}}(\mu)$, 
\begin{equation} 
 \gamma_{ji,\tilde{g}} (\mu) = 
   \frac{\alpha_s (\mu)}{4 \pi}      \, \gamma_{ji,\tilde{g}}^{0}
 + \frac{\alpha_s^2(\mu)}{(4 \pi)^2} \, \gamma_{ji,\tilde{g}}^{1}
 + .....
 \,,
\label{anomaldecomp}
\end{equation}
is possible thanks to the choice of including appropriate powers of 
$g_s(\mu)$ into the definition of the operators 
${\cal O}_{i,\tilde{g}}$, as discussed in Sec.~\ref{ordersplit}. 
Since no NLO results are presented in this paper,
the symbol $\gamma_{ji,\tilde{g}} (\mu)$ will be used in the following 
to indicate the LO quantity $\gamma_{ji,\tilde{g}}^0 (\mu)$.
Similarly the Wilson coefficients $C_{i,\tilde{g}}$ will be indicating
$C_{i,\tilde{g}}^0 $, as already understood in the previous sections.
The indices $i,j$ in~(\ref{coeffdecomp}) and~(\ref{anomaldecomp})
run over all gluino-induced operators: 12 magnetic and chromomagnetic 
operators and 5 times (one for each flavour $q$) 20 four-quark
operators. The anomalous-dimension matrix $ \gamma_{ji,\tilde{g}}$ 
is then a $112\times 112$ matrix. 
It turns out, however, that primed and non-primed operators 
do not mix. This reduces the problem to the evaluation of two
identical $56\times 56$ matrices. 

Moreover, given their lower dimensionality, the dimension-five 
operators 
${\cal O}_{7\tilde{g},\tilde{g}}$, ${\cal O}_{8\tilde{g},\tilde{g}}$,
and ${\cal O}_{7\tilde{g},\tilde{g}}^\prime$,
${\cal O}_{8\tilde{g},\tilde{g}}^\prime$, do not mix with 
dimension-six magnetic operators. 
The $4\times 4$ submatrix for these operators is a block-diagonal 
matrix with $2\times 2$ blocks. The block corresponding to
${\cal O}_{7\tilde{g},\tilde{g}}$, ${\cal O}_{8\tilde{g},\tilde{g}}$
is: 
\begin{equation}
\gamma_{ji,\tilde{g}} = \left[   \begin{array}{cc}
\phantom{-}   18                     &  0        \\[1.1ex]  
 -\displaystyle{\frac{32}{9}}        & \displaystyle{\frac{50}{3}} 
                                 \end{array} \right]
\hspace*{1truecm} (i,j = 7\tilde{g},8\tilde{g})\,,
\label{gammadimfive}
\end{equation}
and differs from the known mixing matrix of the SM operators
${\cal O}_{7}$ and ${\cal O}_{8}$ just by anomalous dimensions of 
the explicit mass ${\overline m}_b$ and of the coupling $g_s^2$ in the 
definition of the operators.

In general, the structure of the remaining $54\times 54$ matrix, corresponding
to the four-quark operators
${\cal O}_{i,\tilde{g}}^{q}$ ${(i=11,...,20;\, q=u,d,c,s,b)}$,
magnetic operators
${\cal O}_{7b,\tilde{g}}$, ${\cal O}_{7c,\tilde{g}}$,  
and the  chromomagnetic operators
${\cal O}_{8b,\tilde{g}}$, ${\cal O}_{8c,\tilde{g}}$, 
is rather complicated. 
The fact that in a LO calculation only the
coefficients $C_{7b,\tilde{g}}$ and 
$C_{7c,\tilde{g}}$ (and corresponding primed coefficients) are
needed at the low scale $\mu_b$, however, simplifies 
the analysis considerably.
Among the four-quark operators, only those with scalar/tensor
Lorentz structure, i.e. 
${\cal O}_{i,\tilde{g}}^{q}$  ${(i=15,...,20)}$,
mix into the magnetic and chromomagnetic operators at order
$\alpha_s$. The vector operators
(${\cal O}_{i,\tilde{g}}^{q}$  ${(i=11,...,14)}$) 
on the other hand mix
neither  into the magnetic and chromomagnetic operators nor into the
scalar/tensor four-quark operators. (The scalar/tensor operators,
however, mix into the vector four-quark operators.)
This implies that the presence
of the four-quark operators with vector structure is completely irrelevant
for the evolution of the coefficients of the magnetic operators. 
The observation that
the scalar/tensor operators with the label $q$ mix into 
${\cal O}_{7q,\tilde{g}}$ and 
${\cal O}_{8q,\tilde{g}}$, with the same $q$, 
together with the fact that scalar/tensor
operators mix among themselves in a flavour-diagonal way,
further simplifies the situation. 
It is indeed possible to restrict the problem at the LO level
to the 
calculation of two $8 \times 8$ matrices, i.e. the two matrices 
corresponding to the operators
${\cal O}_{15,\tilde{g}}^q, {\cal O}_{16,\tilde{g}}^q$,
${\cal O}_{17,\tilde{g}}^q, {\cal O}_{18,\tilde{g}}^q$,
${\cal O}_{19,\tilde{g}}^q,{\cal O}_{20,\tilde{g}}^q$,
${\cal O}_{7q,\tilde{g}},{\cal O}_{8q,\tilde{g}}$,
for $q=b$ and $q=c$.

For the case $q=b$, the result of such a calculation, in which 
the anomalous dimensions due to the explicit powers of the
coupling $\alpha_s$ are again included, is:
\begin{equation}
  \left\{\gamma_{ji,\tilde{g}} \right\}      = \left[
\begin{array}{rrrrrr|rr}
  \displaystyle{\frac{44}{3}}    
&      0        
&      0                         
&      0    
& \displaystyle{\frac{1}{3}}     
&     -1    \ 
& -\displaystyle{\frac{1}{3}}    
&      1    
\\[1.2ex]
      -6                         
& \displaystyle{\frac{98}{3}}   
&      0                         
&      0          
& -\displaystyle{\frac{1}{2}}    
& -\displaystyle{\frac{7}{6}} \  
&     -1                         
&      0         
\\ 
       0                         
&      0     
& \phantom{-} \displaystyle{\frac{44}{3}}     
&      0         
&      0       
&      0                      \
&      0        
&      0        
\\ 
       0       
&      0     
&     -6       
& \phantom{-} \displaystyle{\frac{98}{3}}      
&      0       
&      0                      \
&      0       
&      0        
\\ 
      16      
&    -48    
&      0       
&      0        
& \phantom{-} 36      
&      0                      \   
& \displaystyle{\frac{28}{3}}     
&     -4       
\\[1.3ex]
     -24     
&    -56    
&      0        
&      0        
&      6       
& \phantom{-} 18              \
& \displaystyle{\frac{20}{3}}      
&     -8       
\\[1.2ex] \hline & & & & & & & \\[-1.95ex]
       0       
&      0     
&      0       
&      0        
&      0       
&      0                      \  
&     26       
&      0       
\\[1.1ex] 
       0       
&      0     
&      0       
&      0                      
&      0       
&      0                      \
& -\displaystyle{\frac{32}{9}}     
&\phantom{-} \displaystyle{\frac{74}{3}}       
\end{array} \right] \,. 
\label{gamma0b}
\end{equation}

The anomalous-dimension matrix corresponding to the case $q=c$
differs from the previous one 
in the submatrix responsible for mixing of the four-quark operators 
into the magnetic and chromomagnetic operators:  
 \begin{equation}
  \left\{\gamma_{ji,\tilde{g}} \right\}  = \left[
\begin{array}{rrrrrr|rr}
  \displaystyle{\frac{44}{3}}     
&      0        
&      0          
&      0    
& \displaystyle{\frac{1}{3}}     
&     -1                      \ 
&      0
&      0
\\[1.2ex]
      -6       
& \displaystyle{\frac{98}{3}}   
&      0       
&      0          
& -\displaystyle{\frac{1}{2}}     
& -\displaystyle{\frac{7}{6}} \  
&      0
&      0         
\\
       0       
&      0     
& \phantom{-} \displaystyle{\frac{44}{3}}     
&      0         
&      0       
&      0                      \
&      0        
&      0        
\\ 
       0       
&      0     
&     -6       
& \phantom{-} \displaystyle{\frac{98}{3}}      
&      0       
&      0                      \
&      0       
&      0        
\\[1.1ex] 
      16      
&    -48    
&      0       
&      0        
& \phantom{-} 36      
&      0                      \   
&    -16
&      0
\\[1.1ex]
     -24     
&    -56    
&      0        
&      0        
&      6       
& \phantom{-} 18              \
& -\displaystyle{\frac{16}{3}}      
&     -8       
\\[1.2ex] \hline & & & & & & & \\[-1.95ex]
       0       
&      0     
&      0       
&      0        
&      0       
&      0                      \  
&     26       
&      0       
\\[1.1ex] 
       0       
&      0     
&      0       
&      0                      
&      0       
&      0                      \
& -\displaystyle{\frac{32}{9}}     
&\phantom{-} \displaystyle{\frac{74}{3}}       
\end{array} \right] \,. 
\label{gamma0c}
\end{equation}

Using the anomalous dimensions matrices~(\ref{gammadimfive}),
(\ref{gamma0b}) and~(\ref{gamma0c}),  
the renormalization
group equation~(\ref{RGE}) can be solved by the standard procedure, 
described, for example, in Ref.~\cite{BBL}, using the Wilson coefficients
$C_{i,\tilde{g}}(\mu_W)$ given in Sec.~\ref{WilsonCoeff} as
initial conditions.  The integration of~(\ref{RGE}) for 
$C_{7\tilde{g},\tilde{g}}$ and $C_{8\tilde{g},\tilde{g}}$ 
yields the following expressions for these Wilson coefficients 
at the low scale $\mu_b$:
\begin{eqnarray}
{C_{7\tilde{g},\tilde{g}}(\mu_b)}   & = &  
 \eta^{\frac {27}{23}}
 \,{C_{7\tilde{g},\tilde{g}}}(\mu_W) + 
 {\displaystyle \frac {8}{3}} 
 \left( \eta^{\frac {25}{23}} -\eta^{\frac {27}{23}} \right)
 \,{C_{8\tilde{g},\tilde{g}}}(\mu_W)\,,   
\nonumber \\
{C_{8\tilde{g},\tilde{g}}(\mu_b)}   & = &  
 \eta^{\frac {25}{23}}
 \,{C_{8\tilde{g},\tilde{g}}}(\mu_W)\,. 
\label{evoldimfive}
\end{eqnarray} 
Here and in the following, $\eta$ denotes the ratio 
$\alpha_s(\mu_W)/\alpha_s(\mu_b)$. The low-scale Wilson coefficients 
for the corresponding primed operators are obtained by  replacing 
in~(\ref{evoldimfive}) all the unprimed coefficients with  
primed ones. The same holds for the following coefficients.

The Wilson coefficients of the dimension-six operators 
$C_{7b,\tilde{g}}$ and $C_{8b,\tilde{g}}$  
are at low scale:
\begin{eqnarray}
{C_{7b,\tilde{g}}(\mu_b)}           & = &  
  \eta^{\frac {39}{23}} 
 \,{C_{7b,\tilde{g}}}(\mu_W) + 
{\displaystyle \frac {8}{3}}  
\left( \eta^{\frac {37}{23}} -\eta^{\frac {39}{23}}\right)
 \,{C_{8b,\tilde{g}}}(\mu_W) + {R_{7b,\tilde{g}}}(\mu_b)\,, 
\nonumber \\ 
{C_{8b,\tilde{g}}(\mu_b)}          & = &  
 \eta^{\frac {37}{23}}
 \,{C_{8b,\tilde{g}}}(\mu_W) + {R_{8b,\tilde{g}}}(\mu_b)\,.
\label{evoldimsixb}
\end{eqnarray} 
The remainder functions 
$R_{7b,\tilde{g}}(\mu_b)$ and $R_{8b,\tilde{g}}(\mu_b)$ 
are given in Appendix~\ref{remainders}.  They turn out to be
numerically very small with respect to the other terms on the right-hand
sides of~(\ref{evoldimsixb}).  Notice that, in the approximation
$R_{7b,\tilde{g}}(\mu_b)=R_{8b,\tilde{g}}(\mu_b)=0$, the 
low-scale coefficients 
$C_{7b,\tilde{g}}(\mu_b)$ and 
$C_{8b,\tilde{g}}(\mu_b)$ are simply obtained through the 
integration of~(\ref{RGE}) with the anomalous dimension matrix 
$\gamma_{ji,\tilde{g}}$ reduced to the $2\times 2$ block
of~(\ref{gamma0b})
corresponding to the operators ${\cal O}_{7b,\tilde{g}}$ and 
${\cal O}_{8b,\tilde{g}}$.

Finally, the coefficients 
$C_{7c,\tilde{g}}(\mu_b)$ and $C_{8c,\tilde{g}}(\mu_b)$ 
formally have the same expression as  
$C_{7b,\tilde{g}}(\mu_b)$ and $C_{8b,\tilde{g}}(\mu_b)$,
when the indices $7b$ and $8b$ are replaced by 
$7c$ and $8c$. 
Also in this case, the functions 
$R_{7c,\tilde{g}}(\mu_b)$ and $R_{8c,\tilde{g}}(\mu_b)$, 
listed in Appendix~\ref{remainders}, are numerically small.  
In the approximation 
$R_{7c,\tilde{g}}(\mu_b)=R_{8c,\tilde{g}}(\mu_b)=0$, the 
coefficients $C_{7c,\tilde{g}}(\mu_b)$ and 
$C_{8c,\tilde{g}}(\mu_b)$ vanish identically, since 
the corresponding Wilson coefficients at the matching 
scale are vanishing.

\section{Branching ratio}
\label{branching}

The branching ratio 
$ {\rm BR}(\bar{B}\to X_s\gamma)$ can be expressed as 
\begin{equation}  
 {\rm BR}(\bar{B}\to X_s\gamma) =
 \frac{\Gamma(b \to s \gamma)}{\Gamma_{SL}} \, {\rm BR}_{SL} \,,
\label{bratio}
\end{equation}  
where ${\rm BR}_{SL}=(10.49 \pm 0.46)\%$ is 
the measured semileptonic branching
ratio. To the relevant order in $\alpha_s$, the semileptonic decay width
is given by:
\begin{equation}
\Gamma_{SL} =
 \frac{m_b^5 \, G_F^2 \, |V_{cb}|^2}{192 \pi^3}
 \, g\left(\frac{m_c^2}{m_b^2}\right) \,,
\end{equation}
where the phase-space function $g(z)$ is 
$g(z) = 1 - 8z + 8 z^3 - z^4 - 12 z^2 \log z$.
The decay width for $b \to s \gamma$ reads:
\begin{equation}
 \Gamma(b \to s \gamma) = 
 \frac{m_b^5 \, G_F^2 \, |V_{tb} V_{ts}^*|^2
 \, \alpha}{32 \pi^4} \ 
 \left(\left\vert \hat{C}_7  \right\vert^2 + 
       \left\vert \hat{C}'_7 \right\vert^2
 \right) \,,
\end{equation}
where $\hat{C}_7$ and $\hat{C}'_7$ can be expressed in terms 
of the SM and gluino-induced Wilson coefficients evolved down 
to the decay scale $\mu_b$ as:
\begin{eqnarray}
\hat{C}_7 & = &
 - \frac{16 \sqrt{2} \pi^3 \alpha_s(\mu_b)} {G_F \, V_{tb}V^*_{ts}} 
\left[           C_{7b,\tilde{g}}(\mu_b) + 
 \frac{1}{m_b}   C_{7\tilde{g},\tilde{g}}(\mu_b) + 
 \frac{m_c}{m_b} C_{7c,\tilde{g}}(\mu_b)      
\right] +  C_7(\mu_b)
 \nonumber  \\ 
\hat{C}'_7 & = &
 - \frac{16 \sqrt{2} \pi^3 \alpha_s(\mu_b)} {G_F \, V_{tb}V^*_{ts}} 
\left[           C'_{7b,\tilde{g}}(\mu_b) + 
 \frac{1}{m_b}   C'_{7\tilde{g},\tilde{g}}(\mu_b) +
 \frac{m_c}{m_b} C'_{7c,\tilde{g}}(\mu_b)    
\right] \,.
\label{c7hat}
\end{eqnarray}
Notice that, at the leading logarithmic level, it is not possible to
distinguish between the pole masses $m_b$ and $m_c$ from the
corresponding running quantities at the scale $m_b$ or $m_c$. In the
following, these mass parameters are always treated as pole masses.

\section{Numerical Results}
\label{constraints}

\begin{figure}[p]
\begin{center}
\leavevmode
\epsfxsize= 11.5 truecm
\epsfbox[18 167 580 580]{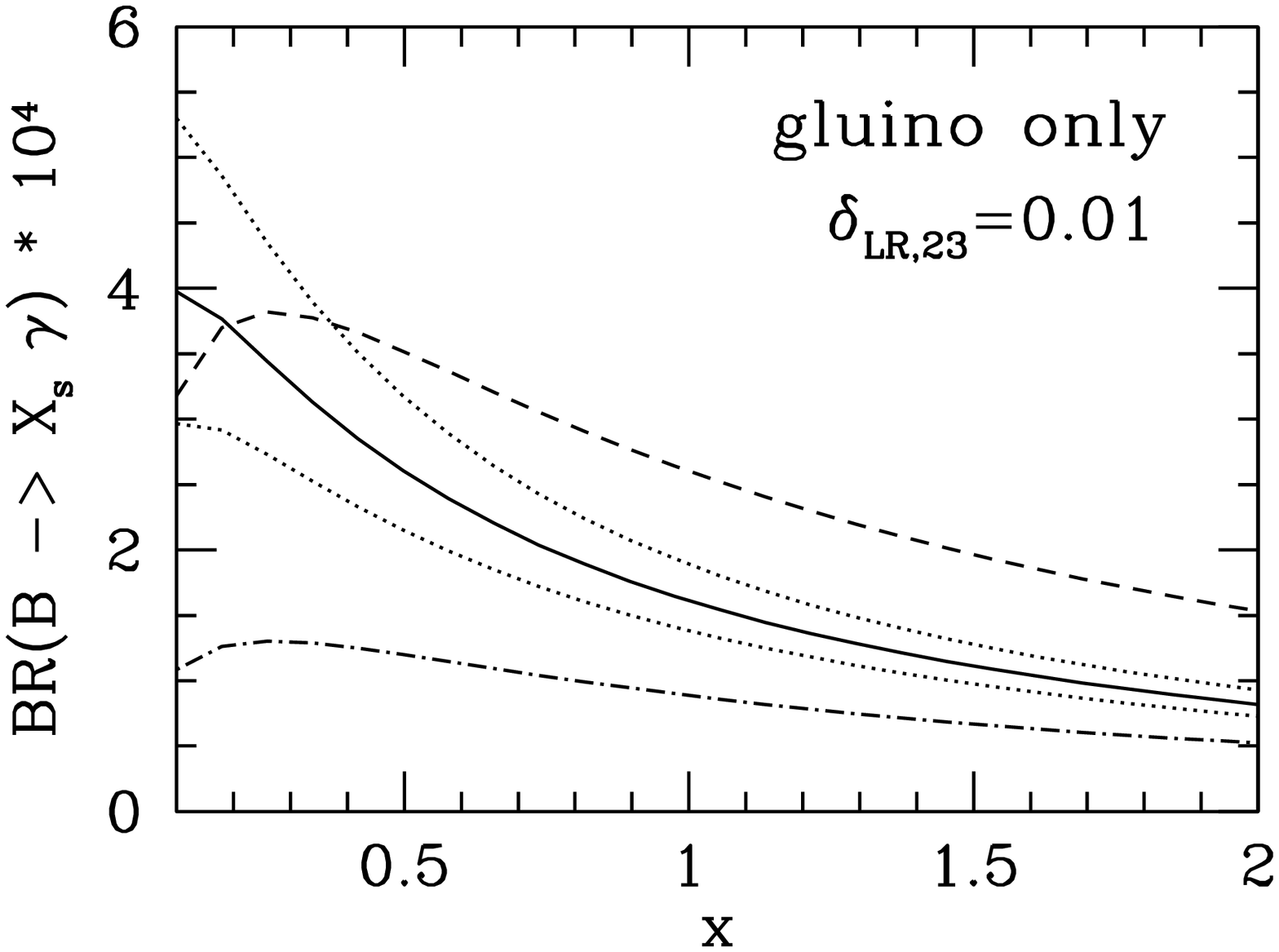}
\end{center}
\caption[f1]{Gluino-induced branching ratio
 ${\rm BR}(\bar{B}\to X_s\gamma)$ as a function of
 $x= m^2_{\tilde{g}}/m^2_{\tilde{q}}$, obtained when the only source
 of flavour violation is $\delta_{LR,23}$ (see text), fixed to the
 value $0.01$, for $m_{\tilde{q}}=500\,$GeV. The solid line shows the
 branching ratio at the LO in QCD, for $\mu_b =4.8\,$GeV and
 $\mu_W = M_W$; the two dotted lines indicate the range of variation
 of the branching ratio when $\mu_b$ spans the interval
 $2.4$--$9.6\,$GeV. Also shown are the values of
 ${\rm BR}(\bar{B}\to X_s\gamma)$ when no QCD corrections are included
 and the explicit factor $\alpha_s(\mu)$ in the gluino-induced
 operators is evaluated at $4.8\,$GeV (dashed line) or at $M_W$
 (dot-dashed line).}
\label{sizeqcd23lr}
\begin{center}
\leavevmode
\epsfxsize= 11.5 truecm
\epsfbox[18 167 580 580]{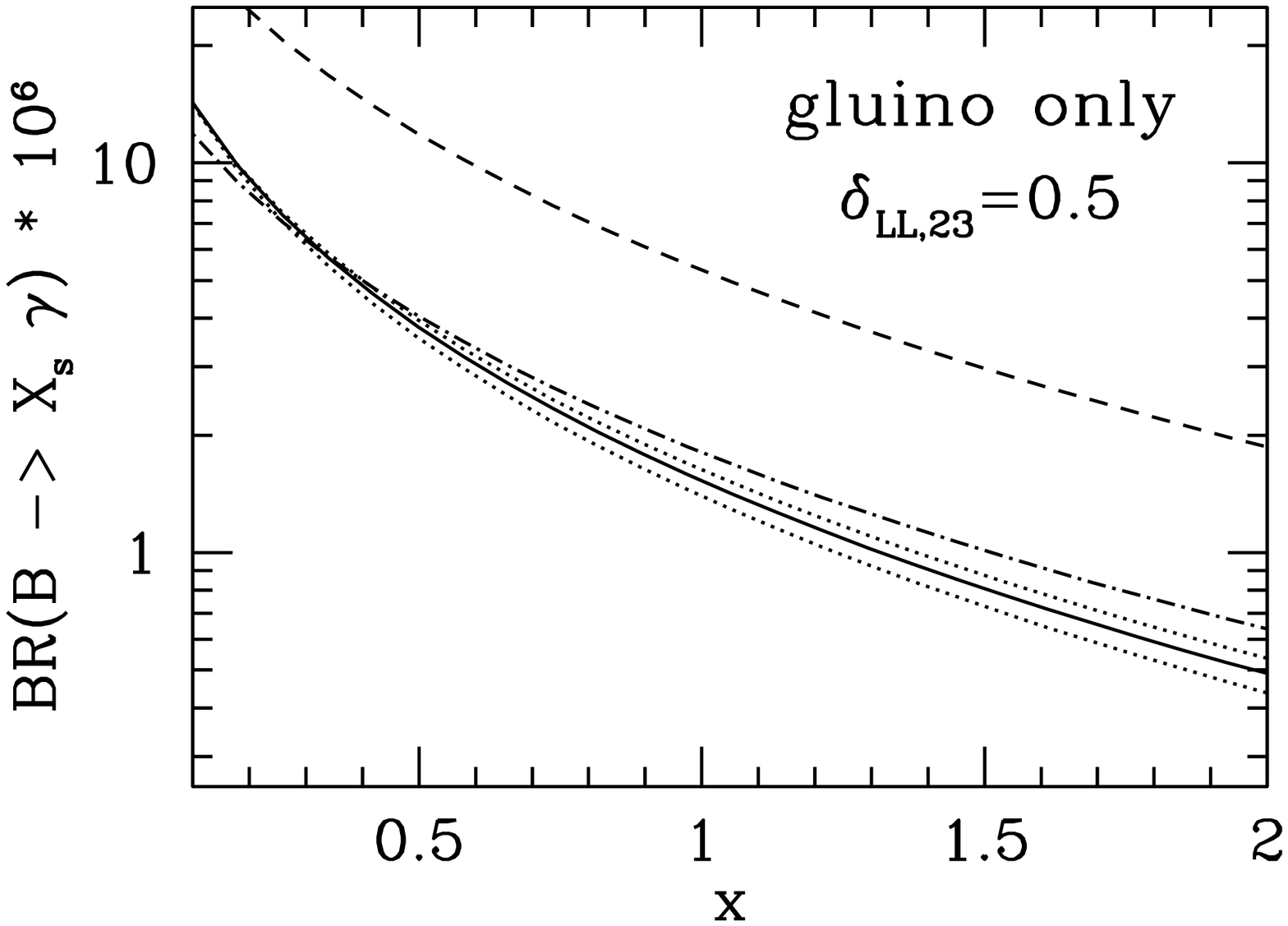}
\end{center}
\caption[f1]{Same as in Fig.~\ref{sizeqcd23lr} when only
 $\delta_{LL,23}$ is non-vanishing and fixed to the value $0.5$.}
\label{sizeqcd23ll}
\end{figure}

Numerical predictions for the QCD-corrected branching ratio 
${\rm BR}(\bar{B}\to X_s\gamma)$ induced by gluino--squark exchange can
be obtained from eqs.~(\ref{bratio})--(\ref{c7hat}).  To show these
results, it is convenient to select one possible source of flavour
violation in the squark sector at a time and assume that all the
remaining ones are vanishing.

Following Ref.~\cite{GGMS}, all diagonal entries in 
$m^2_{\,d,\,LL}$, $m^2_{\,d,\,RR}$, and 
$m^2_{\,u,\,RR}$
are set to be equal and their common value is denoted by
$m_{\tilde{q}}^2$.  The branching ratio can then be studied as a
function of only one off-diagonal element in $m^2_{\,d,\,LL}$ and
$m^2_{\,d,\,RR}$, normalized to $m_{\tilde{q}}^2$, i.e. as a function
of one of the elements
\begin{equation} 
\delta_{LL,ij} = \frac{(m^2_{\,d,\,LL})_{ij}}{m^2_{\tilde{q}}}\,, 
\hspace{1.0truecm}
\delta_{RR,ij} = \frac{(m^2_{\,d,\,RR})_{ij}}{m^2_{\tilde{q}}}\,, 
\hspace{1.0truecm}
(i \ne j) 
\label{deltadefa}
\end{equation}
and/or of one diagonal or off-diagonal element of the $3 \times 3$
matrices $m^2_{\,d,\,LR}$, $m^2_{\,d,\,RL}$ again normalized to 
$m_{\tilde{q}}^2$:
\begin{equation} 
\delta_{LR,ij} = \frac{(m^2_{\,d,\,LR})_{ij}}{m^2_{\tilde{q}}}\,,
\hspace{1.0truecm}
\delta_{RL,ij} = \frac{(m^2_{\,d,\,RL})^\dagger_{ij}}{m^2_{\tilde{q}}}\,.
\phantom{\hspace{1.0truecm}
(i \ne j)} 
\label{deltadefb}
\end{equation}
The corresponding off-diagonal entries in the up-squark 
mass matrix squared,
relevant for the contributions coming from the gluino-induced four-quark
operators~(\ref{penguinboxop}) and~(\ref{boxop}) are set to be equal 
to those in the down-squark mass matrix squared. 
Among the four-quark operators, 
only the scalar/tensor operators~(\ref{boxop}) contribute to 
${\rm BR}(\bar{B}\to X_s \gamma)$, at the LO order in QCD.
Their effect is negligible and the above restriction is not likely to
produce an unnatural reduction of their contribution.
Indeed, due to their
proportionality to $\Gamma^{kb}_{DR} \Gamma^{*ks}_{DL}$, the 
operators ${\cal O}^{q}_{i,\tilde{g}}$ ($i=15,...,20$) are 
generated always together with 
${\cal O}_{7\tilde{g},\tilde{g}}$ and 
${\cal O}_{8\tilde{g},\tilde{g}}$.  As will 
be discussed later, the latter are the numerically important operators 
and the corrections induced e.g. by 
${\cal O}_{8\tilde{g},\tilde{g}}$ on the Wilson coefficent 
$C_{7\tilde{g},\tilde{g}}(\mu_b)$ 
completely overshadow 
the effect of the four-quark operators 
${\cal O}^{q}_{i,\tilde{g}}$. These induce 
corrections of the Wilson coefficient 
$C_{7\tilde{b},\tilde{g}}(\mu_b)$ of the numerically less 
relevant operator
${\cal O}_{7\tilde{b},\tilde{g}}$ that are generically
suppressed by a factor 
 $(m_b/m_{\widetilde{g}})$ at the amplitude level. Analogously,   
the primed scalar/tensor operators 
${\cal O}^{q\,\prime}_{i,\tilde{g}}$ ($i=15,...,20$) 
are also expected to have a very small impact on the decay amplitude.
The vector four-quark operators, on the other hand, can
be generated without the simultaneous generation of 
${\cal O}_{7\tilde{g},\tilde{g}}$ and
${\cal O}_{8\tilde{g},\tilde{g}}$ and no suppression
factor $(m_b/m_{\widetilde{g}})$ is present in this case. 
Therefore, the vector four-quark operators, although entering at NLO only,
are in general expected to
have a larger impact on the decay amplitude than the 
scalar/tensor four-quark operators.
In the context of a NLO analysis, one should 
actually check if the assumption of equal off-diagonal entries in 
the up- and down-squark mass matrices squared 
is not an oversimplification,
affecting the generality of the numerical results.

As for the remaining entries in the squark mass matrices squared, 
the $D$-terms
are calculated using $M_Z=91.18\,$GeV, $\sin^2 \theta_W = 0.2316$,
and $\tan \beta =2$; the $F$-terms $F_{f\,LL}$ and $F_{f\,RR}$, using
$m_b=3\,$GeV and $m_t=175\,$GeV, in the approximation of vanishing
lighter quark masses, whereas $F_{f\,LR}=F_{f\,RL}= 0$ is assumed. It
is obvious that all the information gained through the numerical
evaluation of ${\rm BR}(\bar{B}\to X_s\gamma)$ on the size of
$(m_{\,d,\,LR}^2)_{33}$ can be extended to the combination
$(m_{\,d,\,LR}^2+F_{f\,LR})_{33}$ and
$((m_{\,d,\,LR}^{2})^{\dagger}+F_{f\,RL})_{33}$ in realistic cases, in
which $\mu \ne 0$.
For the diagonal entry $m_{\tilde{q}}^2$, the value 
$m_{\tilde{q}}= 500\,$GeV is in general used. Moreover, it is 
imposed that the eigenvalues of the two $6\times 6$ up- and 
down-squark mass matrices are larger than $150\,$GeV for all 
values of the $\delta$-ratios scanned.
The value of $150\,$GeV is here taken as an average
model-independent lower limit on squark masses, which 
can be inferred from direct searches of squarks at 
hadron colliders.

Finally, the remaining parameter needed to determine the 
branching ratio is: 
\begin{equation}
 x = \frac{m^2_{\tilde{g}}}{m^2_{\tilde{q}}} \,,
\end{equation}
where $m_{\tilde{g}}$ is the gluino mass.

In the following, the SM contribution to 
${\rm BR}(\bar{B}\to X_s\gamma)$ is, in general, added to the gluino
contribution: possible constraints on the flavour-violating sources in
the squark sector should be extracted, keeping into account that the SM
contribution already successfully saturates the experimental result for 
this branching ratio~\cite{ALEPH,CLEO}.
As already stressed in Sec.~\ref{intro}, this analysis applies to 
particular directions of the supersymmetric parameter space, in 
which charged Higgs, chargino and neutralino contributions can be 
safely neglected with respect to the gluino and SM contributions. 
Moreover, it should also be mentioned that the bounds discussed in 
this section on $\delta_{LL,23}$, $\delta_{RR,23}$, $\delta_{LR,23}$, 
and $\delta_{RL,23}$, obtained in these particular directions of 
parameter space, have to be understood in 
an indicative sense, since they are extracted ignoring the error 
of the theoretical calculation.

It is useful to isolate the gluino contribution when 
illustrating the impact of the LO QCD corrections on the gluino-induced
Hamiltonian. In Figs.~\ref{sizeqcd23lr} and~\ref{sizeqcd23ll},
indicated by solid lines, are shown the values of the QCD-corrected
branching ratio obtained, respectively, when only $\delta_{LR,23}$ and
$\delta_{LL,23}$ are non-vanishing.  Their values are fixed in the two
Figures as follows: $\delta_{LR,23}=0.01$ and
$\delta_{LL,23}=0.5$. The branching ratio is plotted as a function of
$x$, i.e. as a function of the gluino mass, for a given value of
$m_{\tilde{q}}$, $m_{\tilde{q}}=500\,$GeV.  Also shown is 
the range of variation
of the branching ratio, delimited by dotted lines, obtained when the
low-energy scale $\mu_b$ spans the interval $2.4$--$9.6\,$GeV. The
matching scale $\mu_W$ is here fixed to $M_W$. As can be seen, the
theoretical estimate of ${\rm BR}(\bar{B}\to X_s\gamma)$ is still largely
uncertain ($\sim \pm 25\%$). An extraction of bounds on 
$\delta_{LL,23}$ and $\delta_{LR,23}$ 
more precise than just an order of magnitude would require, therefore,
the inclusion of NLO QCD corrections. It should be
noticed, however, that the inclusion of corrections at the LO has
already removed the large ambiguity on the value to be assigned to the
factor $\alpha_s(\mu)$ in the gluino-induced magnetic
operators~(\ref{gmagnopb})--(\ref{gmagnopc}). Before adding QCD
corrections, the scale in this factor can assume all values from
$\mu_b$ to $\mu_W$. The corresponding values for 
${\rm BR}(\bar{B}\to X_s\gamma)$ for the two extreme choices of $\mu$
are indicated in Figs.~\ref{sizeqcd23lr} and~\ref{sizeqcd23ll} by
the dot-dashed lines ($\mu=M_W$) and the dashed lines
($\mu=4.8\,$GeV): the branching ratio is virtually unknown!  The
choice $\mu = M_W$ gives values for the non-QCD-corrected 
${\rm BR}(\bar{B}\to X_s\gamma)$ relatively close to the 
band obtained for the QCD-corrected result, in the case shown in 
Fig.~\ref{sizeqcd23ll}, when only $\delta_{LL,23}$ is
non-vanishing. Finding a corresponding value of $\mu$ that minimizes
the QCD corrections in the case studied in Fig.~\ref{sizeqcd23lr},
when only $\delta_{LR,23}$ is different from zero,
depends strongly on the value of $x$.

\begin{figure}[p]
\begin{center}
\leavevmode
\epsfxsize= 11.5 truecm
\epsfbox[18 167 580 580]{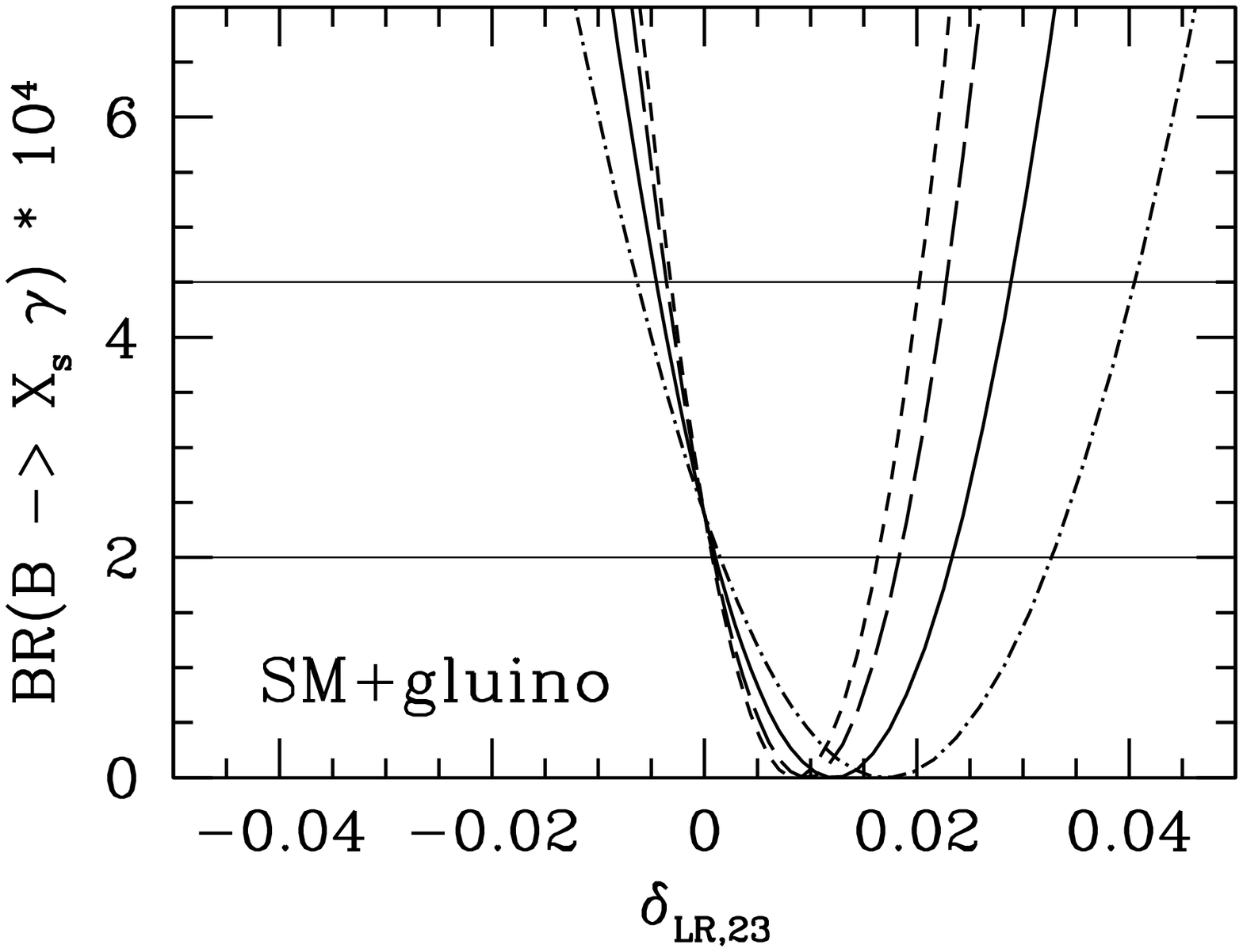}
\end{center}
\caption[f1]{Dependence of the QCD-corrected branching ratio 
 ${\rm BR}(\bar{B}\to X_s\gamma)$, obtained from the SM and gluino
 contributions, on the parameter $\delta_{LR,23}$, when
 $(m_{\,d,\,LR}^2)_{23}$ is the
 only non-vanishing off-diagonal element in the down-squark mass
 matrix squared. The branching ratio is shown for different values of
 $x=m^2_{\tilde{g}}/m^2_{\tilde{q}}$, with $m_{\tilde{q}}=500\,$GeV:
 0.3 (short-dashed line), 0.5 (long-dashed line), 1 (solid
 line), and 2 (dot-dashed line).  Low and matching scales are: 
 $\mu_b= 4.8\,$GeV and $\mu_W=M_W$.}
\label{glsm23lr}
\begin{center}
\leavevmode
\epsfxsize= 11.5 truecm
\epsfbox[18 167 580 580]{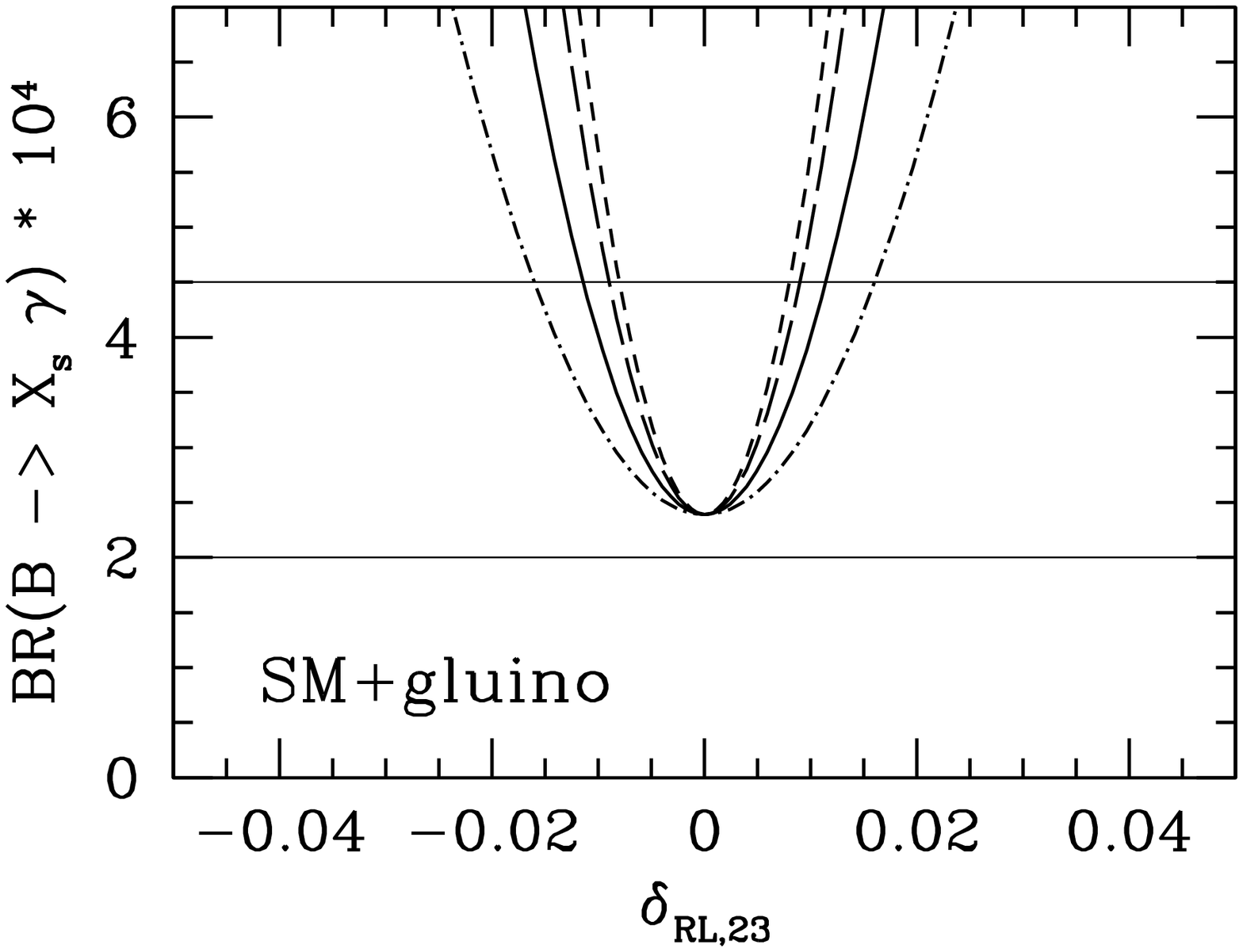}
\end{center}
\caption[f1]{Same as in Fig.~\ref{glsm23lr}, when 
 $\delta_{RL,23}$ is the only source of flavour violation for the
 gluino contribution. The parameter $x$ is fixed to: 0.3 (short-dashed
 line), 0.5 (long-dashed line), 1 (solid line), 2 (dot-dashed line).}
\label{glsm23rl}
\end{figure}

The results in Figs.~\ref{sizeqcd23lr} and~\ref{sizeqcd23ll} also  show
that the operator ${\cal O}_{7b,\tilde{g}}$ gives much smaller
contributions to ${\rm BR}(\bar{B}\to X_s\gamma)$ than the operator
${\cal O}_{7\tilde{g},\tilde{g}}$. Indeed, the branching ratio obtained
through ${\cal O}_{7b,\tilde{g}}$ only is typically suppressed by a 
factor
$(m_b/m_{\tilde{g}})^2$, with respect to that obtained from 
${\cal O}_{7\tilde{g},\tilde{g}}$, 
if similar values of $\delta_{LL,23}$ and
$\delta_{LR,23}$ are chosen.  Analogous considerations hold for 
${\cal O}_{7b,\tilde{g}}^\prime$ and 
${\cal O}_{7\tilde{g},\tilde{g}}^\prime$. The elements
$\delta_{LR,23}$ and $\delta_{RL,23}$ are therefore expected to be the
flavour-violating parameters most efficiently constrained by the
measurement of ${\rm BR}(\bar{B}\to X_s\gamma)$.

In Fig.~\ref{glsm23lr}, the dependence of 
${\rm BR}(\bar{B}\to X_s\gamma)$ is shown as a function of
$\delta_{LR,23}$ when this is the only flavour-violating source.
The two horizontal lines correspond to the minimum and
maximum values, $2 \times 10^{-4}$ and $4.5 \times 10^{-4}$, allowed by
the CLEO measurement.  The branching ratio is obtained by adding the
SM and the gluino contribution calculated for different choices of
$x$, and a fixed value of $m_{\tilde{q}}$: $m_{\tilde{q}}=500\,$GeV.
The values of the gluino mass corresponding to the choices $x=0.3$,
$0.5$, $1$, $2$ are: $m_{\tilde{g}} = 274$, $354$, $500$, 
$707\,$GeV.  The branching ratio is plotted in this Figure for fixed
values of the two scales: $\mu_b = 4.8\,$GeV and $\mu_W = M_W$.  The
gluino contribution interferes constructively with the SM for negative
values of $\delta_{LR,23}$, which are then more sharply constrained
than the positive values. Overall, this parameter cannot exceed the
per cent level. No interference with the SM is present 
when $\delta_{RL,23}$ is the only source of flavour violation, 
as shown in Fig.~\ref{glsm23rl}. The results obtained for 
${\rm BR}(\bar{B}\to X_s\gamma)$ are then symmetric around  
$\delta_{RL,23}=0$ and the constraints on $\vert \delta_{RL,23}\vert$ 
are upper bounds on its absolute value: there are no small values of
$\delta_{RL,23}$ for which the total branching ratio falls off the
band allowed by the CLEO measurement.

\begin{figure}[p]
\begin{center}
\leavevmode
\epsfxsize= 11.5 truecm
\epsfbox[18 167 580 580]{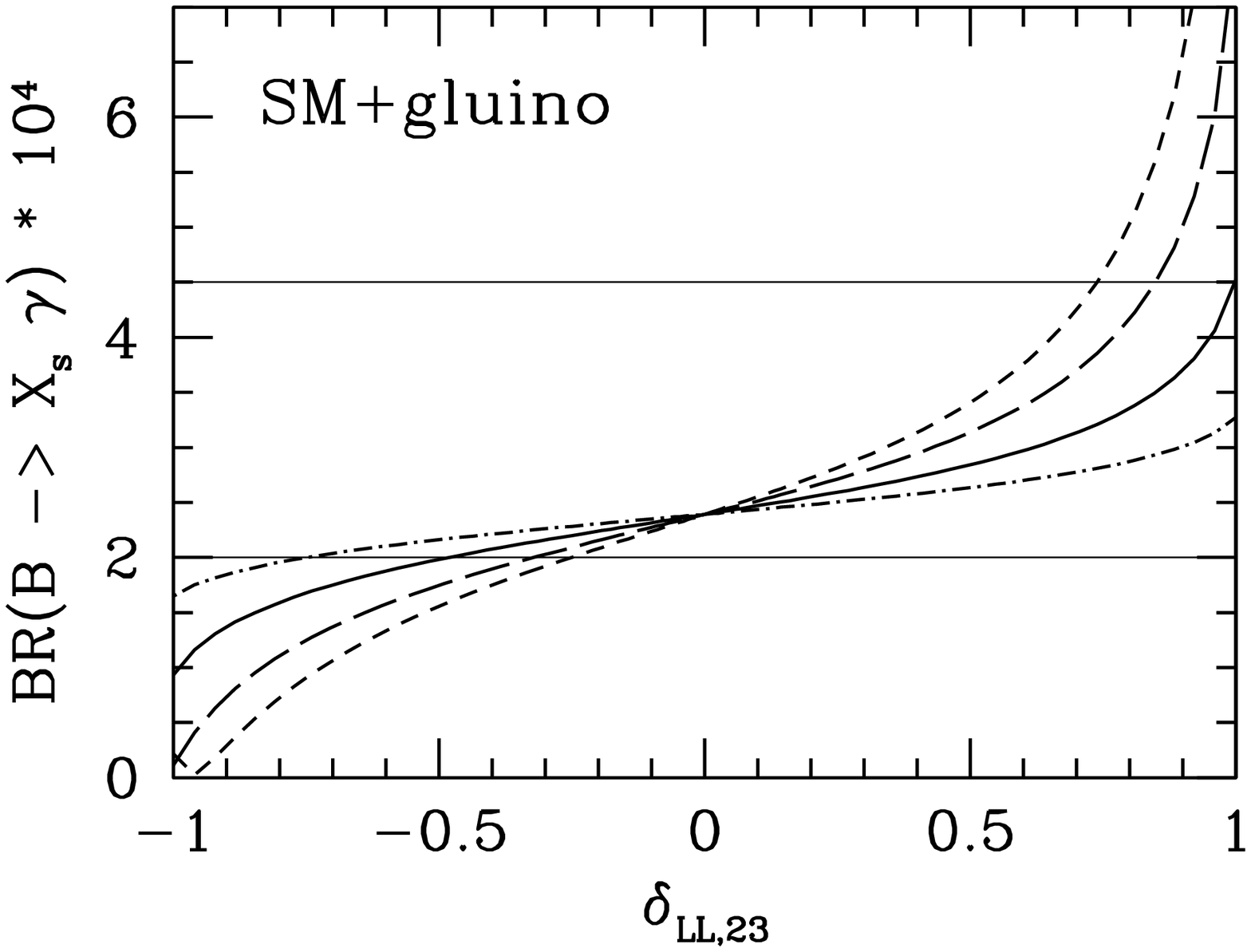}
\end{center}
 \caption[f1]{Same as in Fig.~\ref{glsm23lr}, when 
  $\delta_{LL,23}$ is the only source of flavour violation for the
  gluino contribution. The different lines correspond to: $x=0.3$ 
  (short-dashed line), 0.5 (long-dashed line), 1 (solid line), 2 
  (dot-dashed line).}
 \label{glsm23ll}
\begin{center}
\leavevmode
\epsfxsize= 11.5 truecm
\epsfbox[18 167 580 580]{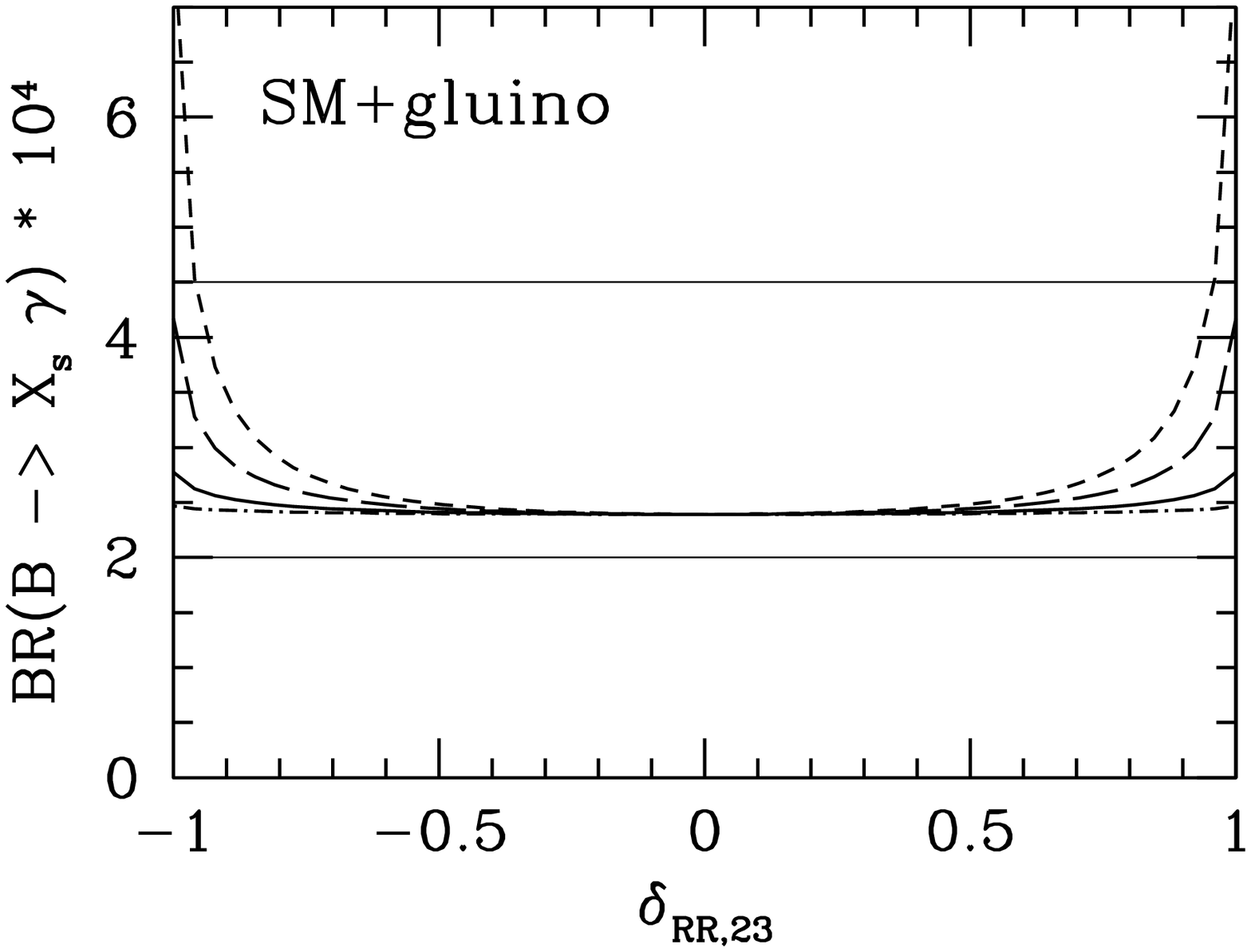}
\end{center}
 \caption[f1]{Same as in Fig.~\ref{glsm23lr}, when 
  $\delta_{RR,23}$ is the only source of flavour violation for the
  gluino contribution. The values of $x$ corresponding to the 
  different lines are: 0.3 (short-dashed
  line), 0.5 (long-dashed line), 1 (solid line), 2 (dot-dashed line).}
 \label{glsm23rr}
 \end{figure}

\begin{figure}[t]
\begin{center}
\leavevmode
\epsfxsize= 11.5 truecm
\epsfbox[18 167 580 580]{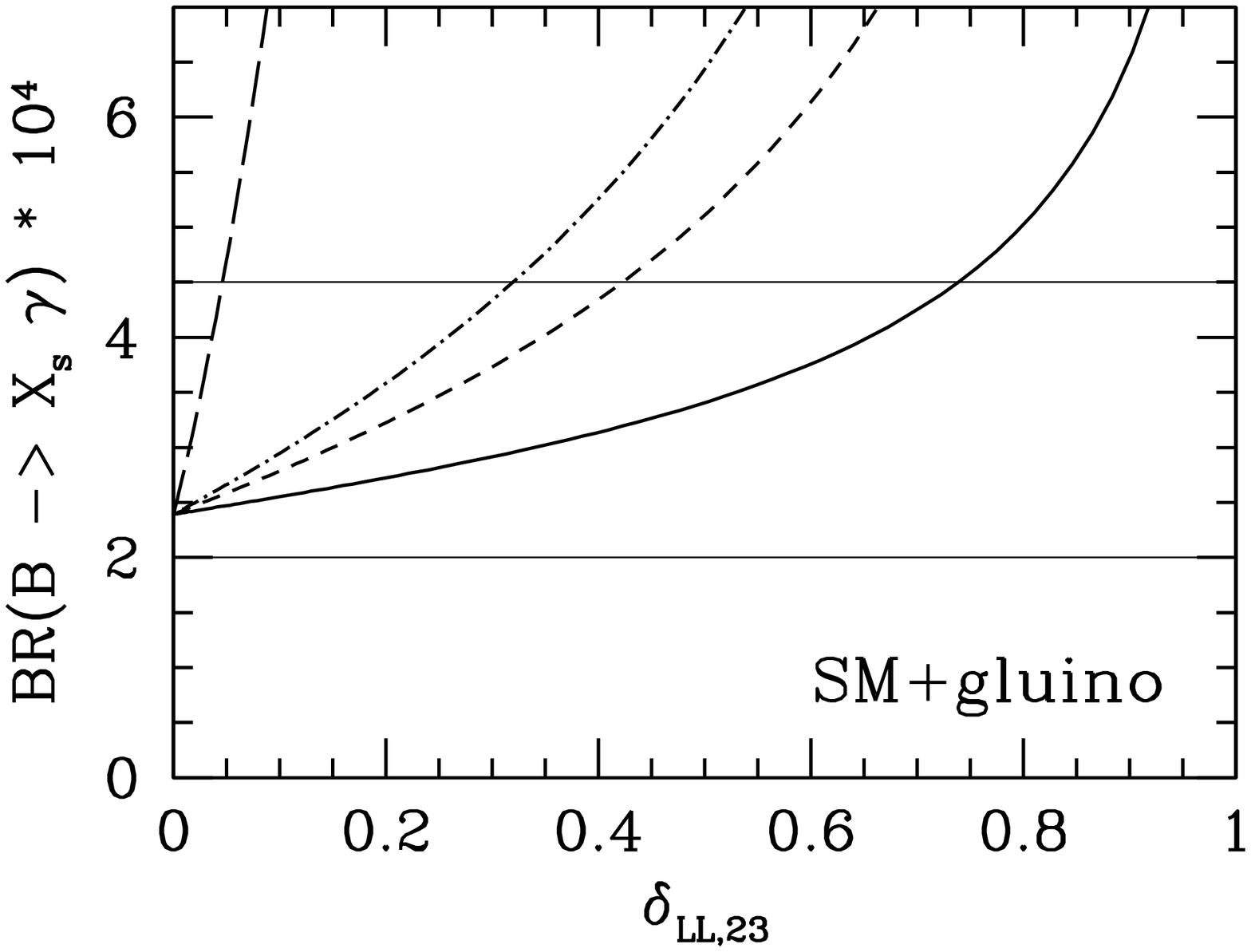}
\end{center}
\caption[f1]{ ${\rm BR}(\bar{B}\to X_s\gamma)$ vs. $\delta_{LL,23}$, 
 when $\delta_{LL,23}$ and
 $\delta_{LR,33}$ are the only sources of chiral-flavour violation.
 The dependence on $\delta_{LL,23}$ is shown for different values of
 $\delta_{LR,33}$: 0 (solid line), 0.006 (short-dashed line), 0.01
 (dot-dashed line), 0.1 (long-dashed line). The value of 
 $x= m^2_{\tilde{g}}/m^2_{\tilde{q}}$ is fixed to $0.3$ and 
 $m_{\tilde{q}}$ to $500\,$GeV.}
\label{glsm23llnew}
\end{figure}

Much weaker is the dependence of ${\rm BR}(\bar{B}\to X_s\gamma)$ on
$\delta_{LL,23}$ if $(m^2_{\,d,\,LL})_{23}$ is the only off-diagonal
element in the down squark mass matrix squared. This dependence is 
illustrated
in Fig.~\ref{glsm23ll} for different choices of $x$ and
$m_{\tilde{q}}=500\,$GeV. The gluino--squark loop generates in this
case only the dimension-six operator ${\cal O}_{7b,\tilde{g}}$ and the
gluino contribution interferes constructively with the SM
contribution for positive $\delta_{LL,23}$.  Notice that 
the mass insertion approximation,  
given the
large values of $\delta_{LL,23}$ allowed by the experimental
measurement, cannot be used in this
case to obtain a reliable estimate of ${\rm BR}(\bar{B}\to X_s\gamma)$, 
whereas it is an excellent approximation of the complete calculation
in the cases shown in Figs.~\ref{glsm23lr} and~\ref{glsm23rl}.
For completeness, also the case in which the only 
off-diagonal element in the down-squark mass matrix squared 
is in the right--right sector, $(m^2_{\,d,\,RR})_{23}\ne 0$,  
is shown in Fig.~\ref{glsm23rr}. The inclusive branching ratio, 
plotted versus the relevant parameter $\delta_{RR,23}$, is now
obtained from the incoherent sum of the SM and gluino contributions and
shows conspicuous deviation from the SM result only for very large
values of $\delta_{RR,23}$.

As already observed, among the operators ${\cal O}_{7b,\tilde{g}}$ and
${\cal O}_{7\tilde{g},\tilde{g}}$, the second one has the stronger
impact on ${\rm BR}(\bar{B}\to X_s\gamma)$. It is then legitimate to
question whether ${\cal O}_{7\tilde{g},\tilde{g}}$ may not provide a
stronger constraint on $\delta_{LL,23}$. Since
${\cal O}_{7\tilde{g},\tilde{g}}$ requires a chirality flip within the
loop, then at least an additional off-diagonal element different from
zero is needed in the left--left sector of the 
down-squark mass matrix squared. Indeed, the
flavour-conserving left-right mixing term $(m^2_{\,d,\,LR})_{33}$ 
together with $(m^2_{\,d,\,LL})_{23}$ can also generate the operator
${\cal O}_{7\tilde{g},\tilde{g}}$; see the first diagram in
Fig.~\ref{glopgtwoins}.  The corresponding branching ratio is shown in
Fig.~\ref{glsm23llnew}, as a function of $\delta_{LL,23}$ for different
choices of $\delta_{LR,33}$. The value of the diagonal entries in the 
squark mass matrix is $m_{\tilde{q}}=500\,$GeV and $m_{\widetilde{g}}$ 
is determined by the choice $x=0.3$. As in the previous plots, low
and matching scales are fixed as $\mu_b = 4.8\,$GeV and $\mu_W = M_W$.
Both parameters $\delta_{LR,33}$ and $\delta_{LL,23}$ are chosen to
be positive. The solid line in this Figure, obtained for
$\delta_{LR,33}=0$, then coincides with the short-dashed line in
Fig.~\ref{glsm23ll}. The SM value of the branching ratio, at the 
LO in QCD, is the value
at which all curves meet for $\delta_{LL,23}=0$. The short-dashed line
is obtained for $(m^2_{\,d,\,LR})_{33}\simeq m_{\tilde{q}} \, m_b$, 
which corresponds to a relatively large trilinear coupling in models in
which the trilinear term in the soft potential is proportional to the
Yukawa couplings.  The corresponding maximally allowed value of
$\delta_{LL,23}$ already is, in this case, considerably smaller than that 
obtained when only the operator ${\cal O}_{7b,\tilde{g}}$ is 
present. Larger values of $\delta_{LR,33}$ obviously induce
even more stringent constraints on $\delta_{LL,23}$.

Two obvious lessons can be learned out of this analysis. First, in
directions of the supersymmetric parameter space in which other
contributions to ${\rm BR}(\bar{B}\to X_s\gamma)$ cannot be neglected,
some of the constraints derived here may be invalidated by
possible interferences among different contributions. An
illustration of this is provided by the comparison of the bounds 
imposed by ${\rm BR}(\bar{B}\to X_s\gamma)$
on $\delta_{LR,23}$ and $\delta_{RL,23}$,
which are different precisely because 
contributions from SM-gluino interferences 
are possible in one case, but not in the
other. The second lesson stems from the observation that different
operators contributing to ${\rm BR}(\bar{B}\to X_s\gamma)$ have very
different numerical relevance. Because of this, it is not necessarily
true that the strongest constraint on a chiral-flavour-violating
sfermion mass term can be derived from the operator that  is generated
by it in the most straightforward way.  Therefore, one cannot but end
this section by stressing again the importance of analyses as complete as
possible, when attempting to use the $b \to s \gamma$ decay as a
model-building tool, constraining the soft supersymmetry-breaking terms.

\section{Summary}
\label{closing}

Gluino-mediated contributions to FCNC processes are useful probes of
chiral-flavour-violating soft breaking terms.  They are in general
cleaner than chargino contributions, which are sensitive also to the
CKM matrix, responsible for flavour violation in the SM and in 2HDMs.
Since they come with a coupling $\alpha_s$, they are usually rather
large. Whether they are indeed much larger than chargino contributions
is a model-dependent issue.

The presence of the coupling $\alpha_s$ makes these contributions also
particularly interesting for FCNC processes in which QCD corrections
play as important a role as the purely electroweak contributions.
Exemplary among these processes is the decay $b \to s \gamma$.  A
specific analysis of the implementation of QCD corrections for the
gluino contribution to this decay is required.  This paper is devoted
to precisely this issue: it shows how to QCD-correct the gluino
contribution to the decay $b \to s \gamma$, using the formalism of
effective Hamiltonian.

It is shown here that, contrary to the common belief, gluino
contributions require an enlargement of the standard basis of
operators needed to describe $b$-$s$ transitions in the SM and 2HDMs.
In the SM, the calculation at
the LO in QCD includes all terms of type $(\alpha_s \log(M_W/m_b))^N$,
whereas the calculation at the NLO resums all terms
$\alpha_s (\alpha_s \log(M^2_W/m^2_b))^N$.  
The program of implementation of QCD corrections in
the SM
requires that at each order in QCD, e.g. LO or NLO,
the anomalous-dimension matrix of the SM operators is calculated at a
higher order in $\alpha_s$ than the matching conditions and the matrix
elements. This is because in the SM, at a certain order in QCD, no
other operator can mix into the magnetic operator 
$ (e/(16 \pi^2))\,(\bar{s} \sigma^{\mu\nu} P_R b)\, F_{\mu\nu}$ 
without the exchange of a virtual gluon. The situation is different in
the case of gluino contributions.  Gluino-induced magnetic operators
acquire corrections as in the SM, when an additional virtual gluon is
exchanged. Moreover, as in the SM, also gluino-induced chromomagnetic
operators mix into the gluino-induced magnetic ones after the on-shell
gluon is connected to a quark line and an additional photon is
radiated. Both operators get first non-vanishing contributions at the
matching scale at order $\alpha_s$, and give QCD-corrected
contributions of type $\alpha_s^2 \log(M^2_W/m^2_b)$.  Gluino-induced 
four-quark operators with first non-vanishing contributions at the
matching scale of ${\cal O}(\alpha_s^2)$, however, can 
mix into the
magnetic operators through the connection of two of the external quark
lines and the emission of an on-shell photon, giving therefore also
corrections of type $\alpha_s^2 \log(M^2_W/m^2_b)$. 
As not all logarithms are due to gluon exchange, their 
systematic resummation is more involved as in the SM.
 
A solution to this problem has been proposed in this paper. The
couplings $\alpha_s$ and $\alpha_s^2$ intrinsically connected with the
gluino exchange are respectively factorized out in the definition of
magnetic and chromomagnetic operators and of operators originating
from box diagrams. With this definition, all gluino-induced operators
are distinguished from the standard set of operators in the effective
Hamiltonian induced by SM and 2HDMs. In particular, the magnetic
operator
${\cal O}_7 = 
  {(e/(16\pi^2))} \,{\overline m}_b \,
 (\bar{s} \sigma^{\mu\nu} P_R b) \, F_{\mu\nu}$
is now distinct from the gluino-induced one 
${\cal O}_{7b,\tilde{g}} = 
 e \,g_s^2 \,{\overline m}_b \,
 (\bar{s} \sigma^{\mu\nu} P_R b) \, F_{\mu\nu}$. 
This in turn has to be distinguished from the lower 
dimensionality operator 
${\cal O}_{7\tilde{g},\tilde{g}} = 
 e \,g_s^2 \,
 (\bar{s} \sigma^{\mu\nu} P_R b) \, F_{\mu\nu}$, 
induced at the matching scale by a loop diagram in which 
chiral-flavour violation is provided, for example, by the 
insertion of a left-right mass term in the squark propagator 
and the insertion of a gluino mass in the gluino propagator. 
Completely new are the four-quark operators, such as 
${\cal O}_{15,\tilde{g}}^q =  
 g_s^4(\mu) (\bar{s} P_L b)\, (\bar{q} P_L q)$, 
with an explicit factor $g_s^4$. 
In total, the inclusion of gluino contributions requires 
$56$ new operators and other additional $56$ with opposite 
chirality. 

With the above definition of gluino-induced operators, an
important goal is achieved. The first non-vanishing contribution to
the gluino-induced Wilson coefficients is of ${\cal O}(\alpha_s^0)$. 
Moreover, the anomalous dimension matrix starts at 
${\cal O}(\alpha_s)$. Consequently,
the integration of the renormalization group equation
yields, 
at first non-vanishing order, terms all of type 
$(\alpha_s \log(M_W/m_b))^N$. The analogy with the LO 
SM contributions is now clear. It is this first non-vanishing 
order that is classified as LO gluino contributions. 
Thus, gluino exchanges induce terms of 
type $\alpha_s (\alpha_s \log(M_W/m_b))^N$ at the LO in QCD, 
to be compared to 
the LO SM contributions of type 
$G_F (\alpha_s \log(M_W/m_b))^N$.
The generalization to the NLO is obvious: it will yield contributions 
$\alpha_s^2 (\alpha_s \log(M_W/m_b))^N$ 
for the gluino-induced operators, versus the contributions 
$G_F \alpha_s (\alpha_s \log(M_W/m_b))^N$ coming from the SM 
set of operators.

A complete LO analysis for the branching ratio of the inclusive decay
$\bar{B}\to X_s\gamma$ coming from SM and gluino-induced contributions
is presented in this paper.  The full anomalous-dimension matrix for
gluino-induced operators is calculated and a simple expression for the
branching ratio is given. The gluino-induced Wilson coefficients are
also listed. They are obtained from the evaluation of one-loop
diagrams mediated by the exchange of gluino and squarks. The mass
eigenstate formalism is adopted as the most suitable for
supersymmetric models with different sources of flavour violation and
with a priori large flavour-violating mass terms.

A numerical analysis for the inclusive branching ratio 
${\rm BR}(\bar{B}\to X_s\gamma)$ due to SM and gluino-induced
contributions is presented.  The QCD corrections to the gluino-induced
contributions are found to be even more crucial than in the SM case.
The non-corrected contributions to the inclusive decay 
$\bar{B}\to X_s\gamma$, in fact, suffer from a severe source of
uncertainty that has no counterpart in the SM. At the zeroth order in
QCD, there is no prescription to fix the scale of the overall factor
$\alpha_s^2$ in the final expression of the branching ratio,
intrinsically due to gluino exchanges: it can range from the matching
scale $\sim M_W$ to the low-scale $\sim m_b$.  Once QCD corrections
are added, the bulk of this ambiguity is removed: this factor of
$\alpha_s^2$ has to be evaluated at a low scale of ${\cal O}(m_b)$,
although the exact value of this scale remains unknown. A similar
uncertainty is due to the fact that the matching scale is only known
to be of ${\cal O}(M_W)$.  Thus, the LO branching ratio still suffers
from matching- and low-scale uncertainties similar in size to those in
the SM results.

Finally, we conclude by recalling that this analysis is valid in
particular directions of the supersymmetric parameter space, in which
charged Higgs, chargino and neutralino contributions can be neglected.
In spite of the still large theoretical error, it provides bounds on
the different sources of flavour violation that are present in these
directions of parameter space. Further studies are called for to
include NLO contributions as well as all the remaining supersymmetric
contributions.

\acknowledgements
\noindent 
We thank G.~Buchalla, P.~Minkowski, and A. Pomarol
for discussions. This work was partially supported by 
 the Schweizerischer Nationalfonds and by the European Commission 
through the TMR Network under contract No. ERBFMX-CT960090.
TH acknowledges financial support by the DOE under grant no
DE-FG03-92-ER 40701 during a visit of the theory group at CALTECH
where part of this work was done.

\newpage 
\appendix 

\section{Functions}
\label{functions}

Listed below are the loop functions 
appearing in the coefficients~(\ref{phgl}) and~(\ref{glgl}):
\begin{eqnarray}
 F_1(x) &\quad = \quad &  \frac{1}{ 12\, (\!x-1)^4}
  \left( x^3 -6x^2 +3x +2 +6x\log x\right)  
  \nonumber \\
                                                 &  &  \nonumber \\
 F_2(x) & \quad = \quad & \frac{1}{ 12\, (\!x-1)^4} 
  \left(2x^3 +3x^2 -6x +1 -6x^2\log x\right) 
  \nonumber \\
                                                 &  &  \nonumber \\
 F_3(x) & \quad = \quad & \frac{1}{\phantom{1} 2\, (\!x-1)^3} 
  \left( x^2 -4x +3 +2\log x\right) 
  \nonumber \\
                                                 &  & \nonumber \\
 F_4(x) & \quad = \quad & \frac{1}{ \phantom{1} 2\, (\!x-1)^3}
  \left( x^2 -1 -2x\log x\right)\,;  
\label{loopfunc}
\end{eqnarray}
those originated by the calculation of penguin diagrams
(see coefficients~(\ref{pengcoeff})): 
\begin{eqnarray}
F_5(x) &\quad = \quad &
  \frac{1}{ 36\, (\!x-1)^4}
  \left(7x^3-36x^2+45x-16 +(18x-12)\log x\right)
  \nonumber \\
                                                 &  & \nonumber \\
F_6(x) &\quad = \quad &
  \frac{1}{ 36\, (\!x-1)^4}
  \left(-11x^3+18x^2-9x+2 +6x^3\log x\right)\,;
\label{penguinfunc}
\end{eqnarray}
and finally, the box-diagram functions:
\begin{eqnarray}
F(x,y) &\quad = \quad & 
   -\frac{1}{x-y}\left[\frac{x \log x}{(x-1)^2}
   -\frac{1}{x-1} - (x\rightarrow y)\right]                 
  \nonumber \\
                                                 &  & \nonumber \\
G(x,y) &\quad =\quad & \phantom{-}
   \frac{1}{x-y}\left[\frac{x^2 \log x }{(x-1)^2}
   -\frac{1}{x-1} - (x\rightarrow y)\right]\,.
\label{boxfunc}
\end{eqnarray}

\section{Wilson Coefficient Remainders}
\label{remainders}

The effect of the four-quark operators~(\ref{boxop}) on the 
evolution of the Wilson coefficient relative to the 
magnetic and chromomagnetic 
operators~(\ref{gmagnopb})--(\ref{gmagnopc}) is encoded in the 
remainder functions 
${R_{7q,\tilde{g}}(\mu_b)}$ and ${R_{8q,\tilde{g}}(\mu_b)}$
($q=b,c$) listed below:
\begin{eqnarray}
{R_{7b,\tilde{g}}(\mu_b)} &  = &  
\left(
- {\displaystyle \frac {2353}{33276}} \,(d_1+d_2) 
+ {\displaystyle \frac {34105\,\sqrt{241}}{8019516}} \,(d_1-d_2)
+ {\displaystyle \frac {100}{141}} \,d_3 
- {\displaystyle \frac {67}{118}}  \,d_6 
\right) 
\,{C^{b}_{15,\tilde{g}}}(\mu_W)\mbox{}   
\nonumber \\[1.01ex]
& + & 
\left(
- {\displaystyle \frac {595}{33276}} \,(d_1+d_2) 
- {\displaystyle \frac {27749\,\sqrt{241}}{8019516}} \,(d_1-d_2) 
- {\displaystyle \frac {32}{141}} \,d_3 
+ {\displaystyle \frac {31}{118}} \,d_6 
\right) 
\,{C^{b}_{16,\tilde{g}}}(\mu_W) 
\nonumber \\[1.01ex]
& + & 
\left(
+ {\displaystyle \frac {1181}{2773}} \,(d_1+d_2)\ \, 
+ {\displaystyle \frac {7131\,\sqrt{241}}{668293}} \,(d_1-d_2)\ \, 
- {\displaystyle \frac {48}{47}} \,d_3 \ \,
+ {\displaystyle \frac {10}{59}} \,d_6 
\right)
\,{C^{b}_{19,\tilde{g}}}(\mu_W) 
\nonumber \\[1.01ex]
& + & 
\left(
+ {\displaystyle \frac {1767}{2773}} \,(d_1+d_2)\ \,  
- {\displaystyle \frac {13487\,\sqrt{241}}{668293}} \,(d_1-d_2) \,
- {\displaystyle \frac {224}{47}} \,d_3 
+ {\displaystyle \frac {206}{59}} \,d_6 
\right)
\,{C^{b}_{20,\tilde{g}}}(\mu_W) \,;
\label{erresevenb}
\end{eqnarray}

\begin{eqnarray}
{R_{8b,\tilde{g}}(\mu_b)} &  = &  
\left( 
+ {\displaystyle \frac {391\,\sqrt{241}}{45308}} \,(d_1-d_2) 
- {\displaystyle \frac {25}{188}} \,\left(d_1+d_2-2 d_3\right) 
\right)
\,{C^{b}_{15,\tilde{g}}}(\mu_W) 
\nonumber \\[1.01ex]
& + & 
\left( 
- {\displaystyle \frac {20\,\sqrt{241}}{11327}} \,(d_1-d_2) \ \,
+ {\displaystyle \frac {2}{47}} \,\left(d_1+d_2-2d_3\right) 
\right)
\,{C^{b}_{16,\tilde{g}}}(\mu_W) 
\nonumber \\[1.01ex]
& + & 
\left( 
- {\displaystyle \frac {231\,\sqrt{241}}{11327}} \,(d_1-d_2)
+ {\displaystyle \frac {9}{47}} \,\left(d_1+d_2-2d_3\right) 
\right)
\,{C^{b}_{19,\tilde{g}}}(\mu_W) 
\nonumber \\[1.01ex]
& + & 
\left(
- {\displaystyle \frac {702\,\sqrt{241}}{11327}} \,(d_1-d_2)
+ {\displaystyle \frac {42}{47}} \,\left(d_1+d_2-2d_3\right)
\right)
\,{C^{b}_{20,\tilde{g}}}(\mu_W)  \,;
\label{erreeightb}
\end{eqnarray}

\begin{eqnarray}
{R_{7c,\tilde{g}}(\mu_b)} & = &  
\left(
- {\displaystyle \frac {2375}{33276}} \,(d_1+d_2) 
+ {\displaystyle \frac {39119\,\sqrt{241}}{8019516}}\,(d_1-d_2)
+ {\displaystyle \frac {576}{2773}} \,d_3
\right.
\nonumber\\ & & \mbox{} 
\phantom{(}
\left.
- {\displaystyle \frac {1273}{33276}} \,(d_4+d_5) 
- {\displaystyle \frac {25937\,\sqrt{241}}{8019516}} \,(d_4-d_5) 
+ {\displaystyle \frac {32}{2773}} \,d_6 
\right)
{C^{c}_{15,\tilde{g}}}(\mu_W)
\nonumber\\[1.01ex]
& + & 
\left(
+ {\displaystyle \frac {1747}{33276}} \,(d_1+d_2) 
+ {\displaystyle \frac {7205\,\sqrt{241}}{8019516}} \,(d_1-d_2) \ 
- {\displaystyle \frac {2824}{8319}} \,d_3 
\right.
\nonumber\\ & & \mbox{} 
\phantom{(}
\left.
- {\displaystyle \frac {5267}{33276}} \,(d_4+d_5) 
- {\displaystyle \frac {85147\,\sqrt{241}}{8019516}} \,(d_4-d_5)
+ {\displaystyle \frac {1528}{2773}} \,d_6 
\right)
{C^{c}_{16,\tilde{g}}}(\mu_W)
\nonumber\\[1.01ex]
& + & 
\left( 
- {\displaystyle \frac {373}{2773}}\,(d_1+d_2) \ \,
- {\displaystyle \frac {17843\,\sqrt{241}}{668293}} \,(d_1-d_2)
- {\displaystyle \frac {4800}{2773}} \,d_3 
\right.
\nonumber\\ & & \mbox{} 
\phantom{(}
\left.
- {\displaystyle \frac {3087}{2773}} \,(d_4+d_5) \ \,
- {\displaystyle \frac {48119\,\sqrt{241}}{668293}} \,(d_4-d_5)
+ {\displaystyle \frac {11720}{2773}} \,d_6 
\right)
{C^{c}_{19,\tilde{g}}}(\mu_W)
\nonumber\\[1.01ex]
& + & 
\left(
+ {\displaystyle \frac {1001}{2773}} \,(d_1+d_2) \ \,
- {\displaystyle \frac {28481\,\sqrt{241}}{668293}} \,(d_1-d_2)
- {\displaystyle \frac {7360}{2773}} \,d_3 
\right.
\nonumber\\ & & \mbox{} 
\phantom{(}
\left.
+ {\displaystyle \frac {907}{2773}} \,(d_4+d_5) \ \,
+ {\displaystyle \frac {11091\,\sqrt{241}}{668293}} \,(d_4-d_5)
+ {\displaystyle \frac {3544}{2773}} \,d_6 
\right)
{C^{c}_{20,\tilde{g}}}(\mu_W)  \,;
\label{erresevenc}
\end{eqnarray}

\begin{eqnarray}
{R_{8c,\tilde{g}}(\mu_b)} & = &  
\left( 
- {\displaystyle \frac {25}{376}} \,(d_1+d_2) 
+ {\displaystyle \frac {391\,\sqrt{241}}{90616}} \,(d_1-d_2)
+ {\displaystyle \frac {216}{2773}} \,d_3
\right.
\nonumber\\ & & \mbox{} 
\phantom{(}
\left.
+ {\displaystyle \frac {13}{472}} \,(d_4+d_5) 
- {\displaystyle \frac {73\,\sqrt{241}}{113752}} \,(d_4-d_5)
\phantom{+ {\displaystyle \frac {216}{2773}} \,d_3}\ \ 
\right)
{C^{c}_{15,\tilde{g}}}(\mu_W) \mbox{} 
\nonumber\\[1.01ex] 
& + & 
\left(
{+\displaystyle \frac {1}{47}} \,(d_1+d_2)\ \,
- {\displaystyle \frac {10\,\sqrt{241}}{11327}} \,(d_1-d_2)\ \, 
- {\displaystyle \frac {353}{2773}} \,d_3 
\right.
\nonumber\\ & & \mbox{} 
\phantom{(}
\left.
+ {\displaystyle \frac {5}{118}} \,(d_4+d_5) \,
+ {\displaystyle \frac {20\,\sqrt{241}}{14219}} \,(d_4-d_5)
\phantom{- {\displaystyle \frac {353}{2773}} \,d_3 }\ \ 
\right)
{C^{c}_{16,\tilde{g}}}(\mu_W)
\nonumber\\[1.01ex] 
& + & 
\left(
+ {\displaystyle \frac {9}{94}} \,(d_1+d_2)\ \,
- {\displaystyle \frac {231\,\sqrt{241}}{22654}} \,(d_1-d_2)
- {\displaystyle \frac {1800}{2773}} \,d_3 
\right.
\nonumber\\ & & \mbox{} 
\phantom{(}
\left.
+ {\displaystyle \frac {27}{118}} \,(d_4+d_5) \,
+ {\displaystyle \frac {393\,\sqrt{241}}{28438}} \,(d_4-d_5)
\phantom{- {\displaystyle \frac {1800}{2773}} \,d_3 }\ \ 
\right)
{C^{c}_{19,\tilde{g}}}(\mu_W)
\nonumber\\[1.01ex] 
& + & 
\left(
+ {\displaystyle \frac {21}{47}} \,(d_1+d_2)\ \,
- {\displaystyle \frac {351\,\sqrt{241}}{11327}} \,(d_1-d_2)
- {\displaystyle \frac {2760}{2773}} \,d_3 
\right.
\nonumber\\ & & \mbox{} 
\phantom{(}
\left.
+ {\displaystyle \frac {3}{59}} \,(d_4+d_5)\ \,
- {\displaystyle \frac {153\,\sqrt{241}}{14219}} \,(d_4-d_5)
\phantom{- {\displaystyle \frac {2760}{2773}} \,d_3}\ \ 
\right)
{C^{c}_{20,\tilde{g}}}(\mu_W)  \,;
\label{erreeightc}
\end{eqnarray}
where the factors $d_1$--$d_6$ are given by
\begin{eqnarray}
& & 
 d_1  =  \eta ^{\frac{(47 +\sqrt{241})}{23}};
\hspace*{0.7truecm}
 d_2  =  \eta ^{\frac{(47 -\sqrt{241})}{23}}; 
\hspace*{0.7truecm}
 d_3  =  \eta ^{\frac{37}{23}};
\nonumber \\ 
& & 
 d_4  =  \eta ^{\frac{(29 +\sqrt{241})}{23}};
\hspace*{0.7truecm}
 d_5  =  \eta ^{\frac{(29 -\sqrt{241})}{23}};
\hspace*{0.7truecm}
 d_6  =  \eta ^{\frac{39}{23}}.
\end{eqnarray}
Notice that in~(\ref{erresevenb})--(\ref{erreeightb}) there is no
dependence on 
${C^{b}_{17,\tilde{g}}}(\mu_W)$ and 
${C^{b}_{18,\tilde{g}}}(\mu_W)$, as there is no dependence on 
${C^{c}_{17,\tilde{g}}}(\mu_W)$ and 
${C^{c}_{18,\tilde{g}}}(\mu_W)$
in~(\ref{erresevenc})--(\ref{erreeightc}). By inspecting the 
two anomalous-dimension matrices in eqs.~(\ref{gamma0b}) 
and~(\ref{gamma0c}), it is easy to see that the two 
operators  
${\cal O}_{17,\tilde{g}}^{q}$, 
${\cal O}_{18,\tilde{g}}^{q}$,
do not mix with the remaining ones 
${\cal O}_{15,\tilde{g}}^{q}$, 
${\cal O}_{16,\tilde{g}}^{q}$,
${\cal O}_{19,\tilde{g}}^{q}$, 
${\cal O}_{20,\tilde{g}}^{q}$,
${\cal O}_{7q,\tilde{g}}$, ${\cal O}_{8q,\tilde{g}}$
in either of the two cases, $q=b$ and $q=c$.

\end{document}